# Ergodic Fading One-sided Interference Channels without State Information at Transmitters

Yan Zhu and Dongning Guo
Department of Electrical Engineering and Computer Science
Northwestern University
Evanston, IL 60208, USA

**Abstract**

This work studies the capacity region of a two-user ergodic interference channel with fading, where only one of the users is subject to interference from the other user, and the channel state information (CSI) is only available at the receivers. A layered erasure model with one-sided interference and with arbitrary fading statistics is studied first, whose capacity region is completely determined as a polygon. Each dominant rate pair can be regarded as the outcome of a trade-off between the rate gain of the interference-free user and the rate loss of the other user due to interference. Using insights from the layered erasure model, inner and outer bounds of the capacity region are provided for the one-sided fading Gaussian interference channels. In particular, the inner bound is achieved by artificially creating layers in the signaling of the interference-free user. The outer bound is developed by characterizing a similar trade-off as in the erasure model by taking a "layered" view using the *incremental channel* approach. Furthermore, the gap between the inner and outer bounds is no more than 12.772 bits per channel use per user, regardless of the signal-to-noise ratios and fading statistics.

## I. Introduction

The capacity region of Gaussian interference channels, comprised of one or more interfering links, remains open for more than thirty years. Etkin, Tse and Wang [1] have recently made an important progress by characterizing the capacity region of the two-user Gaussian interference channel to within one bit. Since then, several new results have been obtained, including the sum rate in special interference regimes [2], [3], the degrees of freedom of Gaussian interference channels [4]–[8] and MIMO Gaussian interference channels [9]. The capacity of *fading* interference channels has also been studied, e.g., in [10], [11], where the focus has been on scenarios where channel state information (CSI) is fully available at the transmitters as well as at the receivers.

This work studies fading interference channels where the instantaneous channel state is available at the receivers but not at the transmitters. This is the case in many practical systems where the channel state can only be measured by the receivers, which cannot inform the transmitters of the state accurately in a timely manner through a feedback link. Specifically, this paper assumes independent (fast) fading over time, where the fading *statistics* are known to the transmitters. Note that the result can also be applied to some situations where the transmitters are given an estimate of the channel state over a coding block, but cannot track its instantaneous variations. This study is different than the work of Raja, Prabhakaran, and Viswanath [12] on compound interference channels, where the channel state (from a finite set) is unknown to the transmitters but remains static over the course of a codeword. The key issue therein is to find a coding scheme which is simultaneously compatible with all interference configurations. The results of [12] are applicable to (slow) block fading interference channels with no CSI at transmitters. The current paper, however, investigates the ergodic case where the code is designed to average over all fading states.

To make progress, this paper considers interference channels with two single-antenna users where the interference is *one-sided*, *i.e.*, only one of the users is subject to interference from the other user. Note that the capacity region of such a channel, also known as Z-interference channel, is open even without fading. Such an interference model is suitable if one of the receivers is within the range of both transmitters, while the other receiver is out of the range of the interfering transmitter. One scenario for this case is a linear network of four nodes with information flow to one direction, where every other node transmits, whose range covers both the intended receiver downstream, as well as the unintended receiver upstream.

Like a number of recent works (*e.g.*, [6], [13]–[16]), this paper makes use of the *deterministic model* approach to glean insights to good coding schemes for general interference models. Despite of its simplicity, the deterministic model captures two key physical phenomenons of wireless channels, namely, the broadcast nature of wireless transmission, and the superposition of multiple signals at the receiver. In particular, fading wireless channels can be simplified to a time-varying version of the deterministic model, where the state of a link corresponds to the number of most significant bits or not erased by noise. The capacity region of such a *layered erasure model* for two-user fading broadcast channel has been established by Tse and Yates, who then apply the insight to obtain a constant-gap characterization of the capacity region of the corresponding fading Gaussian broadcast channel [17].

The main contributions of this paper include:

- The exact capacity region of layered erasure one-sided interference channel with arbitrary fading statistics is established.
- Using insights from the converse result for the layered erasure model, a new outer bound for the capacity region of the fading Gaussian one-sided interference channel is obtained.
- A specific coding scheme is shown to achieve a rate region to within a gap of 12.772 bits per channel use per user from the outer bound, regardless of the fading statistics, signal-to-noise ratios (SNR) and interference-to-noise ratios.

It should be noted that recent results on the layered erasure model due to Aggarwal *et al.* are special case of the general result in this paper. These include the capacity region for uniformly very strong interference (Theorem 3 in [18]) and ergodic very strong interference (Theorem 6), and the sum-capacity for uniformly strong but not very strong interference (Theorem 4), uniformly weak interference (Theorem 7) and a special class of mixed interference (Theorem 9).

The remainder of this paper is organized as follows. The Gaussian fading channel with one-sided interference and the corresponding layered erasure channel are described in Section II. The main results in this paper are summarized in Section III. In order to make our development more accessible, the capacity region for single-layer erasure channel is established first in Section IV, and the development for the general case is relegated to Section V. The result for the Gaussian fading model is found in Section VI. Conclusion is drawn in Section VII.

## II. Models and Notation

Consider an interference channel with two pairs of transmitters and receivers, where the message of transmitter 1 is intended to receiver 1, and the message of transmitter 2 is intended to receiver 2. It is assumed that the interference is one-sided from transmitter 2 to receiver 1, so that the direct link of user 2 is free of interference.

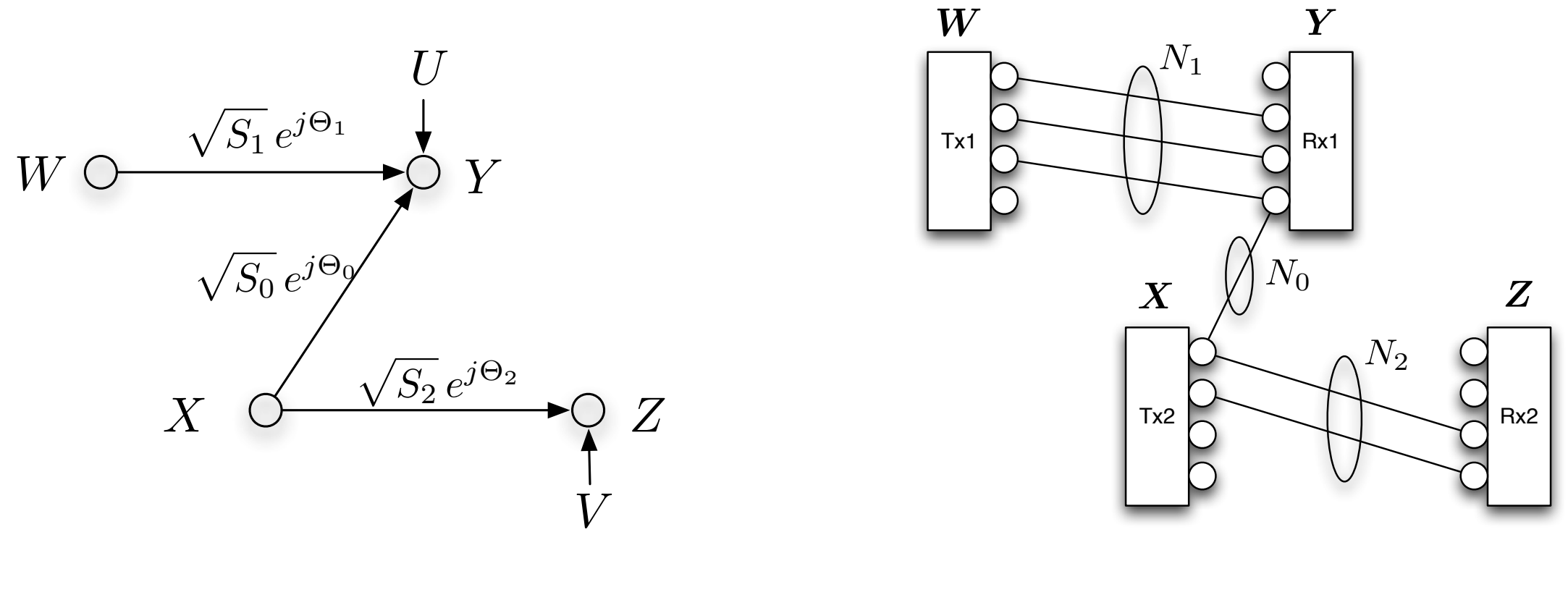

(a) (b)

Fig. 1. One-sided interference channels with fading. (a) A Gaussian model. (b) A layered erasure model.

### *A. The Gaussian Model*

Let $W$, $X$, $Y$ and $Z$ denote the transmitted and received signals of user 1 and user 2, respectively. Consider the following input–output relationship over each time interval $m = 1, \dots, n$:

$$Y_m = \sqrt{S_{1m}}\, e^{j\Theta_{1m}} W_m + \sqrt{S_{0m}}\, e^{j\Theta_{0m}} X_m + U_m \tag{1a}$$

$$Z_m = \sqrt{S_{2m}}\, e^{j\Theta_{2m}} X_m + V_m \tag{1b}$$

where $(S_{1m}, \Theta_{1m})$ and $(S_{2m}, \Theta_{2m})$ denote the channel gain and phase of the two direct links, respectively, and $(S_{0m}, \Theta_{0m})$ denotes the gain and phase of the interference link from transmitter 2 to receiver 1. Such a channel is depicted by Fig. 1(a). For convenience, let the additive noise $\{U_m\}$ and $\{V_m\}$ consist of independent circularly symmetric complex Gaussian (CSCG) variables with unit variance. Let the average power of each transmitted codeword be constrained by 1. The state $S_{im}$ can be regarded as the SNR of the corresponding link. Since CSI is not available at the transmitter, the input signals and the state of the three links are mutually independent. It is assumed that the fading process for each link $\{S_{im}, \Theta_{im}\}$ is independent and identically distributed (i.i.d.) over time $m = 1, \ldots, n$, and that the amplitude process and phase process of each fading process are also mutually independent. Thus all $6n$ variables, $S_{im}, \Theta_{im}$, with $i = 0, 1, 2$ and $m = 1, \ldots, n$, are mutually independent. Furthermore, the phases $\Theta_i$ $(i = 0, 1, 2)$ are assumed to be uniformly distributed on $[0, 2\pi)$. Finally, it is assumed that the fading states are known to the receivers but not to the transmitters.

Note that we often drop the time index when referring to the statistics of an i.i.d. process over time. For example, $S_i$ is identically distributed as $S_{im}$ for $i = 0, 1, 2$.

### *B. The Layered Erasure Model*

In the spirit of the deterministic models introduced in [19], the layered erasure channel model for the one-sided interference channel is depicted by Fig. 1(b) and described as follows. Let the signals emitted by transmitters 1 and 2 at the $m$-th time interval be denoted by $\boldsymbol{W}[m]$ and $\boldsymbol{X}[m]$ respectively, which take values in $\mathbb{F}_2^q$, where $\mathbb{F}_2$ represents the binary Galois field and $\mathbb{F}_2^q$ denotes the $q$-vector space with underlying field $\mathbb{F}_2$. Let $\underline{\boldsymbol{s}}$ denote a $q \times q$ matrix with $s_{i+1,i} = 1$ for all $i = 1, \ldots, q-1$ and all other elements being 0, so that $\underline{\boldsymbol{s}}[x_1, x_2, \ldots, x_q]^{\mathsf{T}} = [0, x_1, \ldots, x_{q-1}]^{\mathsf{T}}$ represents a single shift, and $\underline{\boldsymbol{s}}^n \boldsymbol{X}[m]$ denotes a downward shift of the elements of the vector $\boldsymbol{X}[m]$ with its $n$ least significant bits dropped out and $n$ zeros padded from the top of the vector. The received signals at time interval $m$ are then expressed as:

$$\boldsymbol{Y}[m] = \underline{\boldsymbol{s}}^{q-N_1[m]} \boldsymbol{W}[m] \oplus \underline{\boldsymbol{s}}^{q-N_0[m]} \boldsymbol{X}[m] \tag{2a}$$

$$\boldsymbol{Z}[m] = \underline{\boldsymbol{s}}^{q-N_2[m]} \boldsymbol{X}[m] \tag{2b}$$

where $\{N_0[m]\}$, $\{N_1[m]\}$ and $\{N_2[m]\}$ are integer random processes taking values in $\{0, \ldots, q\}$, which represent the fading states of the three physical links. Let $\{N_0[m]\}$, $\{N_1[m]\}$ and $\{N_2[m]\}$ be mutually independent, and each of the three processes be i.i.d. over time (so that the channel is memoryless). It is further assumed that the fading states are known to both receivers but not to the transmitters.

We introduce the following notation for the layered erasure model for convenience. For a random vector $\boldsymbol{X} \in \mathbb{F}_2^q$, let $X_i$ denote its $i$-th element and $\boldsymbol{X}_i^j$ denote $[X_i, \ldots, X_j]^{\mathsf{T}}$. For a vector process $\boldsymbol{X}[1], \ldots, \boldsymbol{X}[M]$, we use $(X_i)_l^k$ to denote the sequence $X_i[l], \ldots, X_i[k]$, and use $(\boldsymbol{X}_i^j)_l^k$ to denote the sequence $\boldsymbol{X}_i^j[l], \ldots, \boldsymbol{X}_i^j[k]$. The indexes outside the parentheses always refer to time. Binary addition of vectors of different length is aligned at the least significant bits; *e.g.*, if $n_1 \geq n_2$, then $\boldsymbol{X}_1^{n_1} \oplus \boldsymbol{W}_1^{n_2} = [X_1, \ldots, X_{n_1-n_2}, X_{n_1-n_2+1} \oplus W_1, \ldots, X_{n_1} \oplus W_{n_2}]^{\mathsf{T}}$. Since the channel described by (2) is memoryless, we often suppress the time index to describe the model as:

$$\begin{aligned}\boldsymbol{Y} &= \boldsymbol{W}_1^{N_1} \oplus \boldsymbol{X}_1^{N_0}\\ \boldsymbol{Z} &= \boldsymbol{X}_1^{N_2}\,.\end{aligned}$$

Furthermore, the distribution of an i.i.d. sequence of random variables is often represented by the variable with the time index suppressed, *e.g.*, $N_i[m]$ are identically distributed as $N_i$.

## III. Main Results

Throughout the paper, all information units are bits and all logarithms are of base 2.

*Theorem 1:* The capacity region of the one-sided fading Gaussian interference channel (1) is contained in following region:

$$\overline{\mathcal{R}} = \left\{ (R_1, R_2) \;\middle|\; \begin{array}{l} 0 \leq R_1 \leq \mathbb{E}\log\left(1+S_1\right) \\ 0 \leq R_2 \leq \mathbb{E}\log\left(1+S_2\right) \\ R_1 + \omega R_2 \leq 1 + \mathbb{E}\log\left(1+S_1\right) + \omega\mathbb{E}\log\left(1+\frac{S_0}{S_1+1}\right) \\ \qquad\qquad + \int_0^{\infty} (\omega\beta(\gamma)-\alpha(\gamma))^{+}\,\frac{\log e}{1+\gamma}\mathsf{d}\gamma\,, \quad \forall\omega\in[0,1] \end{array} \right\} \tag{3}$$

where

$$\alpha(\gamma) = \mathsf{P}\left(\frac{S_0}{S_1+1} < \gamma \leq S_0\right) \tag{4}$$

and

$$\beta(\gamma) = \mathbb{E}\left[\mathsf{P}\left(S_2 \geq \gamma\right) - \mathsf{P}\left(\frac{S_0}{S_1+1} \geq \gamma \middle| S_1\right)\right]^{+}. \tag{5}$$

Furthermore, regardless of the fading statistics, the outer bound can be achieved to within a gap of at most 12.772 bits/s/Hz per user.

*Theorem 2:* The capacity region of the one-sided layered erasure interference channel (2) is

$$\mathcal{C} = \left\{ (R_1, R_2) \;\middle|\; \begin{array}{l} 0 \leq R_2 \leq \mathbb{E}N_2 \\ 0 \leq R_1 + \omega R_2 \leq \mathbb{E}N_1 + \omega\mathbb{E}\left[N_0 - N_1\right]^{+} + \sum_{l=1}^{q}\left(\omega\beta_l - \alpha_l\right)^{+}, \; \forall\omega\in[0,1] \end{array} \right\} \tag{6}$$

where, for every $l \in \{1, \ldots, q\}$,

$$\alpha_l = \mathsf{P}\left(N_0 - N_1 < l \le N_0\right) \tag{7}$$

and

$$\beta_l = \mathbb{E}\left[\mathsf{P}\left(N_2 \ge l\right) - \mathsf{P}\left(N_0 - N_1 \ge l | N_1\right)\right]^+ . \tag{8}$$

It is clear that the rate region $\overline{\mathcal{R}}$ and the capacity region $\mathcal{C}$ are each surrounded by a collection of simple affine constraints. As we shall see, this is due to the trade-off between the gain in the rate of user 2 and the loss in the rate of user 1 due to interference from user 2, depending on the signaling of the users.

Note that in Theorem 2, setting $\omega = 0$ in the second constraint in (6) yields the single-user bound $0 \le R_1 \le \mathbb{E}N_1$ for the rate of user 1. This however does not apply to Theorem 1 because of the extra constant $1$ in the third constraint of (3).

In the subsequent sections, we first prove Theorem 2 in Sections IV and V. Insights developed from the capacity-achieving scheme for the layered erasure model are then adapted to prove Theorem 1 for the fading Gaussian interference channel in Section VI. Because the proof of Theorem 2 for the general layered erasure channel is still quite involved, we first prove Theorem 2 in the special case of a single layer to illustrate the key ideas and techniques.

We also note that in some special cases, the capacity region or sum-capacity of the fading Gaussian interference channel can be exactly characterized, which, however, are not implied by Theorem 1. Relevant results are given in Section VI-C.

## IV. Proof for the Single-layer Erasure Model

Assume a single layer, *i.e.*, $q = 1$, throughout this section. Denote the erasure probability of the link labeled with $N_i$ as $\epsilon_i$ and let $\bar{\epsilon}_i = 1 - \epsilon_i$ for notational convenience. Evidently, $\bar{\epsilon}_i$ is the probability that the input symbol actually traverses the corresponding link. The region $\mathcal{C}$ defined in Theorem 2 with $q = 1$ is quite simple, as is illustrated in Fig. 2 for all possible configurations of the parameters. The region is precisely described in the following proposition.

*Proposition 1:* Let $q = 1$. In case of strong interference, *i.e.*, $\bar{\epsilon}_0 \ge \bar{\epsilon}_2$, the region $\mathcal{C}$ defined in (6) is the pentagon with boundary constraints $0 \le R_1 \le \bar{\epsilon}_1$, $0 \le R_2 \le \bar{\epsilon}_2$, and

$$R_1 + R_2 \le 1 - \epsilon_0 \epsilon_1 , \tag{9}$$

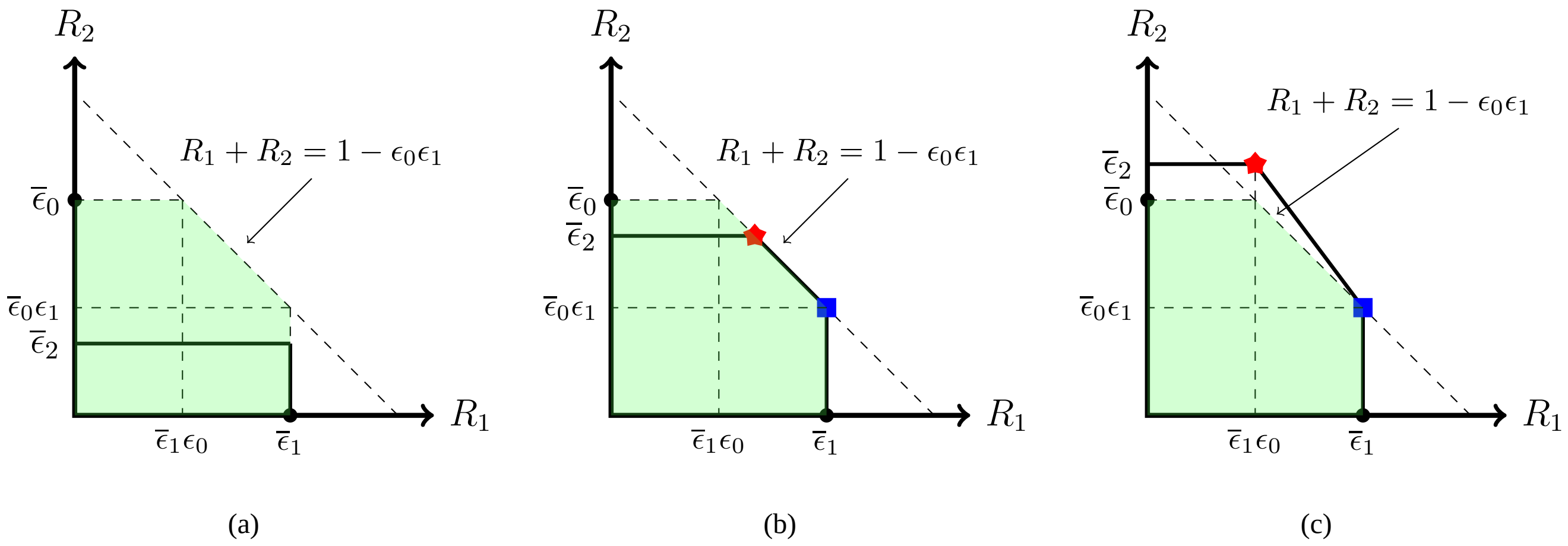

Fig. 2. Capacity region for single-layer erasure channel with different cases drawn by solid lines. (a) $\overline{\epsilon}_2 \le \overline{\epsilon}_0\epsilon_1$. (b) $\overline{\epsilon}_0 \ge \overline{\epsilon}_2 \ge \overline{\epsilon}_0\epsilon_1$. (c) $\overline{\epsilon}_2 \ge \overline{\epsilon}_0$.

which reduces to a rectangle when $\epsilon_2 \le \overline{\epsilon}_0\epsilon_1$. In case of weak interference, *i.e.*, $\overline{\epsilon}_2 \ge \overline{\epsilon}_0$, $\mathcal{C}$ is the pentagon with boundary constraints $0 \le R_1 \le \overline{\epsilon}_1$, $0 \le R_2 \le \overline{\epsilon}_2$, and

$$R_1 + \frac{\alpha_1}{\beta_1} R_2 \le \overline{\epsilon}_1 + \frac{\alpha_1}{\beta_1}\overline{\epsilon}_0\epsilon_1 \tag{10}$$

where $\alpha_1 = \overline{\epsilon}_0\overline{\epsilon}_1$ and $\beta_1 = \overline{\epsilon}_2 - \epsilon_1\overline{\epsilon}_0$.

*Proof:* By (7) and (8),

$$\alpha_1 = \mathsf{P}\left(N_0 = 1, N_1 = 1\right) = \overline{\epsilon}_0\overline{\epsilon}_1$$

and

$$\begin{aligned}\beta_1 &= \mathbb{E}[\mathsf{P}\left(N_2 = 1\right) - \mathsf{P}\left(N_0 = 1, N_1 = 0 | N_1\right)]^+ \\ &= \overline{\epsilon}_1\overline{\epsilon}_2 + \epsilon_1(\overline{\epsilon}_2 - \overline{\epsilon}_0)^+.\end{aligned}$$

If $\overline{\epsilon}_0 \ge \overline{\epsilon}_2$, then $\alpha_1 \ge \beta_1$, so that the second constraint in (6) reduces to $0 \le R_1 + \omega R_2 \le \overline{\epsilon}_1 + \omega\overline{\epsilon}_0\epsilon_1$ for all $\omega \in [0, 1]$. For every $\omega$, the upper bound passes the point $(\overline{\epsilon}_1, \overline{\epsilon}_0\epsilon_1)$, hence the tightest of such bounds is the one with $\omega = 1$, *i.e.*, (9).

If, on the other hand, $\overline{\epsilon}_0 \le \overline{\epsilon}_2$, then $\alpha_1 \le \beta_1 = \overline{\epsilon}_2 - \epsilon_1\overline{\epsilon}_0$, so that the second constraint in (6) becomes

$$R_1 + \omega R_2 \le \overline{\epsilon}_1 + \omega\overline{\epsilon}_0\epsilon_1 + (\omega\beta_1 - \alpha_1)^+ \tag{11}$$

for all $\omega \in [0,1]$. For every $\omega \in [0, \alpha_1/\beta_1]$, the upper bound becomes $R_1 + \omega R_2 \leq \overline{\epsilon}_1 + \omega \overline{\epsilon}_0 \epsilon_1$, the tightest of which is (10) achieved at $\omega = \alpha_1/\beta_1$. For every $\omega \in [\alpha_1/\beta_1, 1]$, the upper bound becomes $R_1 + \omega R_2 \leq \epsilon_0 \overline{\epsilon}_1 + \omega \overline{\epsilon}_2$. It is not difficult to see that all of these bounds as well as (10) pass the point $(\epsilon_0 \overline{\epsilon}_1, \overline{\epsilon}_2)$. Because $R_2 \leq \overline{\epsilon}_2$, the tightest of these bounds is still (10), which describes the dominant face of the region $\mathcal{C}$. ■

In the remainder of this section, it is shown that the region $\mathcal{C}$ described in Proposition 1 is indeed the capacity region. The region is first shown to be achievable, and then a matching converse is established.

### A. *Proof of Achievability*

Since $\boldsymbol{X} = X$ and $\boldsymbol{W} = W$ are scalars, and $N_i = 0$ or $1$, we can denote $\boldsymbol{X}_1^{N_i}$ by $N_i X$ and $\boldsymbol{W}_1^{N_1}$ by $N_1 W$. In each sub-figure of Fig. 2, we shadow the pentagon region enclosed by the axes, the lines $R_1 = \overline{\epsilon}_1$, $R_2 = \overline{\epsilon}_0$, and $R_1 + R_2 = 1 - \epsilon_0 \epsilon_1$, which is the capacity region of the following multiple access channel (MAC):

$$Y = N_1 W \oplus N_0 X \, . \tag{12}$$

Note that if an achievable rate pair $(R_1, R_2)$ for channel (2) falls into the MAC capacity region, then the message from transmitter 2 can be decoded at receiver 1. With these in mind, we investigate the achievability for all two possible cases:

If $\overline{\epsilon}_2 \leq \overline{\epsilon}_0$, $\mathcal{C}$ is contained in the MAC capacity region (see Fig. 2(a) and Fig. 2(b)). Let Ber $(p)$ denote the Bernoulli distribution which puts probability masses of $p$ and $1-p$ at values $1$ and $0$, respectively. Any rate pair in $\mathcal{C}$ can be achieved by using Ber $(1/2)$ inputs and letting receiver 1 decode messages from both transmitters.

If $\overline{\epsilon}_2 \geq \overline{\epsilon}_0$, it suffices to show that the two corner points $(\overline{\epsilon}_1, \overline{\epsilon}_0 \epsilon_1)$ and $(\epsilon_0 \overline{\epsilon}_1, \overline{\epsilon}_2)$, which are marked with star and square in Fig 2(c), respectively, are achievable. Because the point $(\overline{\epsilon}_1, \overline{\epsilon}_0 \epsilon_1)$ is also a corner point of the MAC channel capacity region, it can be achieved. To achieve the second point, both users can use random codebooks generated according to Ber $(1/2)$ distribution. Let the code rate of user 2 be $\overline{\epsilon}_2$. Note that if the fading state $(N_0, N_1) = (0, 1)$, then $Y = W$; for all other realizations of $(N_0, N_1)$, $Y$ is independent of $W$. Therefore, from receiver 1's viewpoint, this is equivalent to an erasure channel with erasure probability $1 - \epsilon_0 \overline{\epsilon}_1$. Thus the rate $\epsilon_0 \overline{\epsilon}_1$ is achievable by user 1, which shows that the corner point $(\epsilon_0 \overline{\epsilon}_1, \overline{\epsilon}_2)$ is also achievable in this case.

To assist the study of general cases, it is helpful to further investigate the achievability of star point in Fig 2(b), whose coordinate is $(\epsilon_2 - \epsilon_0\epsilon_1, \overline{\epsilon}_2)$. In particular, the coding scheme can be constructed explicitly using the rate splitting method [20] as follows. We split user 1 into two virtual users: one encodes its message through random coding with random variable $U \sim \text{Ber}\,(\delta/2)$, and the other encodes its message with random variable $V \sim \text{Ber}\,(1 - 1/(2-\delta))$, where $\delta \in [0,1]$. Let $W = \max(U, V)$. Clearly, $\mathsf{P}\,(W=0) = \mathsf{P}\,(U=0)\,\mathsf{P}\,(V=0) = 1/2$ so that $W[1], \ldots, W[n]$ is an i.i.d $\text{Ber}\,(1/2)$ sequence. Also let user 2 generate its codebook using i.i.d. $\text{Ber}\,(1/2)$ distribution. Receiver 1 can first decode the message encoded in $U$ at rate

$$\begin{aligned}
R_{11}(\delta) &= \mathcal{I}\,(Y; U | N_1, N_0) \\
&= \mathsf{P}\,(N_1 = 1)\,\mathsf{P}\,(N_0 = 0)\,\mathcal{I}\,(W; U) \\
&= \overline{\epsilon}_1 \epsilon_0 (\mathcal{H}\,(W) - \mathcal{H}\,(W|U)) \\
&= \overline{\epsilon}_1 \epsilon_0 \left[1 - \frac{2-\delta}{2}\mathcal{H}\left(\frac{1}{2-\delta}\right)\right] \,,
\end{aligned}$$

then it is able to decode the message of user 2 as long as the rate does not exceed

$$\begin{aligned}
R_2(\delta) &= \mathcal{I}\,(Y; X | U, N_0, N_1) \\
&= \mathsf{P}\,(N_0 = 1)\,\mathsf{P}\,(N_1 = 0)\,\mathcal{H}\,(X) + \mathsf{P}\,(N_0 = N_1 = 1)\,\mathcal{I}\,(W \oplus X; X | U) \\
&= \overline{\epsilon}_0 \epsilon_1 + \overline{\epsilon}_0 \overline{\epsilon}_1 \left[\mathcal{H}\,(W \oplus X | U) - \mathcal{H}\,(W|U)\right] \\
&= \overline{\epsilon}_0 \epsilon_1 + \overline{\epsilon}_0 \overline{\epsilon}_1 \left[1 - \frac{2-\delta}{2}\mathcal{H}\left(\frac{1}{2-\delta}\right)\right] \,,
\end{aligned}$$

and finally decode the message encoded in $V$ at rate

$$\begin{aligned}
R_{12}(\delta) &= \mathcal{I}\,(V; Y | U, X, N_0, N_1) \\
&= \overline{\epsilon}_1 \mathcal{I}\,(V; W | U) \\
&= \overline{\epsilon}_1 \frac{2-\delta}{2}\mathcal{H}\left(\frac{1}{2-\delta}\right) \,.
\end{aligned}$$

Note that the rate for user 1 is $R_1(\delta) = R_{11}(\delta) + R_{12}(\delta)$ and it is easy to verify that $R_1(\delta) + R_2(\delta) = 1 - \epsilon_0\epsilon_1$. Furthermore, as $\delta$ varies from 0 to 1, the point $(R_1, R_2)$ moves from $(\overline{\epsilon}_1, \overline{\epsilon}_0\epsilon_1)$ to $(\overline{\epsilon}_1\epsilon_0, \overline{\epsilon}_0)$. Since $\overline{\epsilon}_0 \geq \overline{\epsilon}_2 \geq \overline{\epsilon}_0\epsilon_1$, there exits some $\delta^*$ such that $\big(R_1(\delta^*), R_2(\delta^*)\big) = (\epsilon_2 - \epsilon_0\epsilon_1, \overline{\epsilon}_2)$. Meanwhile, with i.i.d. $\text{Ber}\,(1/2)$ input $(X)_1^n$, rate $\overline{\epsilon}_2$ can be achieved at receiver 2. Therefore, the corner point $(\epsilon_2 - \epsilon_0\epsilon_1, \overline{\epsilon}_2)$ can be achieved using rate splitting if $\overline{\epsilon}_0 \geq \overline{\epsilon}_2 \geq \overline{\epsilon}_0\epsilon_1$.

### B. Proof of Converse

Every achievable rate pair $(R_1, R_2)$ must satisfy $R_1 \leq \overline{\epsilon}_1$ and $R_2 \leq \overline{\epsilon}_2$. Therefore, it suffices to show that the rate pair must satisfy (9) in the strong-interference case where $\overline{\epsilon}_0 \geq \overline{\epsilon}_2$, and must satisfy (10) in the weak-interference case where $\overline{\epsilon}_0 \leq \overline{\epsilon}_2$.

It is easy to see that, because the two decoders operate independently, the capacity region of the one-sided interference channel depends only on the marginal distribution of the channel outputs conditioned on the inputs, but *not* on the joint conditional distribution [21]. It is assumed in the remainder of this section that the random variables $N_0[m]$ and $N_2[m]$ are "aligned" such that $\mathsf{P}\left(N_0[m] \cdot N_2[m] = 1\right) = \min(\overline{\epsilon}_0, \overline{\epsilon}_2)$ for every $m$, whereas the state variables $\{N_1[m]\}$ remain independent of $\{N_0[m], N_2[m]\}$. Clearly, if the realization of the weaker one between $N_0[m]$ and $N_2[m]$ is equal to 1, then the realization of the stronger one must also be equal to 1. This does not change the capacity region. It is important to note that the alignment does not change $\alpha_1$ and $\beta_1$ either because they depend only on the marginal distributions of $N_0$ and $N_2$.

Consider first the case $\overline{\epsilon}_0 \geq \overline{\epsilon}_2$. For notational simplicity, let $\boldsymbol{N} = (N_0, N_1, N_2)$ so that $(\boldsymbol{N})_1^n$ denotes all fading states from time 1 to time $n$, *i.e.*, $(\boldsymbol{N})_1^n = \{N_i[j] : \ i = 0, 1, 2, \text{and } j = 1, \dots, n\}$. By Fano's inequality, $R_1$ must satisfy

$$\begin{aligned} nR_1 - n\delta_n &\leq \mathcal{I}\left(Y[1], \dots, Y[n]; W[1], \dots, W[n] | (\boldsymbol{N})_1^n\right) \\ &= \mathcal{H}\left((Y)_1^n | (\boldsymbol{N})_1^n\right) - \mathcal{H}\left((Y)_1^n | (W)_1^n, (\boldsymbol{N})_1^n\right) \\ &\leq n(1 - \epsilon_0\epsilon_1) - \mathcal{H}\left((N_0 X)_1^n | (\boldsymbol{N})_1^n\right) \end{aligned} \tag{13}$$

for some $\delta_n$ vanishingly small as $n \to \infty$, where (13) follows from that $\mathcal{H}\left((Y)_1^n | (\boldsymbol{N})_1^n\right)$ is maximized by setting both $(W)_1^n$ and $(X)_1^n$ to be i.i.d Ber$\,(1/2)$ sequence. Also due to Fano's inequality, $R_2$ must satisfy

$$\begin{aligned} nR_2 - n\delta_n &\leq \mathcal{I}\left((Z)_1^n; (X)_1^n | (\boldsymbol{N})_1^n\right) \\ &= \mathcal{H}\left((N_2 X)_1^n | (\boldsymbol{N})_1^n\right) . \end{aligned} \tag{14}$$

Note that $\overline{\epsilon}_0 \geq \overline{\epsilon}_2$ by assumption, so that $N_0 \geq N_2$, and, thus $\mathcal{H}\left((N_0 X)_1^n | (\boldsymbol{N})_1^n\right) \geq \mathcal{H}\left((N_2 X)_1^n | (\boldsymbol{N})_1^n\right)$. Comparing (13) and (14) yields (9) as $n \to \infty$.

Next, consider the case of $\overline{\epsilon}_0 \leq \overline{\epsilon}_2$. Let $(\widetilde{W})_1^n = (\widetilde{W}[1], \dots, \widetilde{W}[n])$ be an i.i.d. Ber$\,(1/2)$ sequence independent of all channel inputs and channel states, and let $\widetilde{Y} = N_1 \widetilde{W} \oplus N_0 X$.

Fano's inequality requires that

$$\begin{aligned}
&nR_1 - n\delta_n \\
&\quad \leq \mathcal{I}\left((Y)_1^n; (W)_1^n | (\boldsymbol{N})_1^n\right) \\
&\quad = \mathcal{H}\left((N_1 W \oplus N_0 X)_1^n | (\boldsymbol{N})_1^n\right) - \mathcal{H}\left((N_0 X)_1^n | (\boldsymbol{N})_1^n\right) \\
&\quad \leq \mathcal{H}\left((N_1 \widetilde{W} \oplus N_1 W \oplus N_0 X)_1^n \middle| (\boldsymbol{N})_1^n\right) - \mathcal{H}\left((N_0 X)_1^n | (\boldsymbol{N})_1^n\right) && (15) \\
&\quad = \mathcal{H}\left((N_1 \widetilde{W} \oplus N_0 X)_1^n \middle| (\boldsymbol{N})_1^n\right) - \mathcal{H}\left((N_0 X)_1^n | (\boldsymbol{N})_1^n\right) && (16) \\
&\quad = \mathcal{I}\left((\widetilde{Y})_1^n; (N_1 \widetilde{W})_1^n \middle| (\boldsymbol{N})_1^n\right)
\end{aligned}$$

where (15) follows from data processing theorem and (16) is due to the fact that $\widetilde{W} \oplus W$ is identically distributed as $\widetilde{W}$. Breaking down the mutual information in another way, we obtain

$$\begin{aligned}
&nR_1 - n\delta_n \\
&\quad \leq \mathcal{H}\left((N_1 \widetilde{W})_1^n \middle| (\boldsymbol{N})_1^n\right) - \mathcal{H}\left((N_1 \widetilde{W})_1^n \middle| (\widetilde{Y})_1^n, (\boldsymbol{N})_1^n\right) \\
&\quad = n\bar{\epsilon}_1 - \mathcal{H}\left((N_1 \widetilde{W})_1^n \middle| (\widetilde{Y})_1^n, (\boldsymbol{N})_1^n\right) \\
&\quad = n\bar{\epsilon}_1 - \mathcal{H}\left((N_0 X)_1^n \middle| (\widetilde{Y})_1^n, (\boldsymbol{N})_1^n\right) . && (17)
\end{aligned}$$

Also by Fano's inequality,

$$\begin{aligned}
&nR_2 - n\delta_n \\
&\quad \leq \mathcal{I}\left((N_2 X)_1^n; (X)_1^n | (\boldsymbol{N})_1^n\right) \\
&\quad \leq \mathcal{I}\left((N_2 X)_1^n, (\widetilde{Y})_1^n; (X)_1^n \middle| (\boldsymbol{N})_1^n\right) \\
&\quad = \mathcal{I}\left((\widetilde{Y})_1^n; (X)_1^n \middle| (\boldsymbol{N})_1^n\right) + \mathcal{I}\left((N_2 X)_1^n; (X)_1^n \middle| (\widetilde{Y})_1^n, (\boldsymbol{N})_1^n\right) \\
&\quad \leq n\bar{\epsilon}_0 \epsilon_1 + \mathcal{H}\left((N_2 X)_1^n \middle| (\widetilde{Y})_1^n, (\boldsymbol{N})_1^n\right) . && (18)
\end{aligned}$$

The upper bounds (17) and (18) can be understood as follows: The rate pair $(\bar{\epsilon}_1, \bar{\epsilon}_0 \epsilon_1)$ can be achieved by letting user 1 decode and cancel the signal of user 2 as shown in Section IV-A. By choosing the signaling $X$, user 2 can improve his/her own rate by $\mathcal{H}\left((N_2 X)_1^n \middle| (\widetilde{Y})_1^n, (\boldsymbol{N})_1^n\right)$ at user 1's rate expense in the amount of $\mathcal{H}\left((N_0 X)_1^n \middle| (\widetilde{Y})_1^n, (\boldsymbol{N})_1^n\right)$ due to interference. In the following, we consider the trade-off between the rate loss of user 1 and the rate gain of user 2 over all choices of the signal $(X)_1^n$. Using Marton-like expansion [22], [23] and the chain rule,

$$\begin{aligned}
&\mathcal{H}\left((N_2X)_1^n \middle| (\widetilde{Y})_1^n, (\boldsymbol{N})_1^n\right) - \mathcal{H}\left((N_0X)_1^n \middle| (\widetilde{Y})_1^n, (\boldsymbol{N})_1^n\right) \\
&= \sum_{i=1}^n \left\{ \mathcal{H}\left((N_2X)_1^i, (N_0X)_{i+1}^n \middle| (\widetilde{Y})_1^n, (\boldsymbol{N})_1^n\right) - \mathcal{H}\left((N_2X)_1^{i-1}, (N_0X)_i^n \middle| (\widetilde{Y})_1^n, (\boldsymbol{N})_1^n\right) \right\} \\
&= \sum_{i=1}^n \left\{ \mathcal{H}\left((N_2X)_1^{i-1}, (N_0X)_{i+1}^n \middle| (\widetilde{Y})_1^n, (\boldsymbol{N})_1^n\right) + \mathcal{H}\left((N_2X)_i \middle| (N_2X)_1^{i-1}, (N_0X)_{i+1}^n, (\widetilde{Y})_1^n, (\boldsymbol{N})_1^n\right) \right. \\
&\quad \left. -\mathcal{H}\left((N_2X)_1^{i-1}, (N_0X)_{i+1}^n \middle| (\widetilde{Y})_1^n, (\boldsymbol{N})_1^n\right) - \mathcal{H}\left((N_0X)_i \middle| (N_2X)_1^{i-1}, (N_0X)_{i+1}^n, (\widetilde{Y})_1^n, (\boldsymbol{N})_1^n\right) \right\} \\
&= \sum_{i=1}^n \left\{ \mathcal{H}\left((N_2X)_i \middle| (N_2X)_1^{i-1}, (N_0X)_{i+1}^n, (\widetilde{Y})_1^n, (\boldsymbol{N})_1^n\right) \right. \\
&\qquad \left. -\mathcal{H}\left((N_0X)_i \middle| (N_2X)_1^{i-1}, (N_0X)_{i+1}^n, (\widetilde{Y})_1^n, (\boldsymbol{N})_1^n\right) \right\} . \qquad (19)
\end{aligned}$$

The expression (19) concerns only the conditional entropy of single random variables. To bound (19), we need following lemma:

*Lemma 1:* Assume alignment of $N_0$ and $N_2$. Let $\boldsymbol{T}$ be a collection of random variables which are independent of $\boldsymbol{N} = (N_0, N_1, N_2)$. Let $\widetilde{W}$ be a Ber$(1/2)$ random variable independent of $X$. Then

$$\frac{1}{\alpha_1}\mathcal{H}\left(N_0X \middle| N_0X \oplus N_1\widetilde{W}, \boldsymbol{T}, \boldsymbol{N}\right) = \mathcal{H}(X|\boldsymbol{T}) = \frac{1}{\beta_1}\mathcal{H}\left(N_2X \middle| N_0X \oplus N_1\widetilde{W}, \boldsymbol{T}, \boldsymbol{N}\right) .$$

*Proof:* Since there is no conditional uncertainty about $N_0X$ unless $N_0 = N_1 = 1$,

$$\begin{aligned}
\mathcal{H}\left(N_0X \middle| N_0X \oplus N_1\widetilde{W}, \boldsymbol{T}, \boldsymbol{N}\right) &= \mathsf{P}(N_1 = 1, N_0 = 1)\,\mathcal{H}(X|\boldsymbol{T}) \\
&= \alpha_1 \mathcal{H}(X|\boldsymbol{T}) .
\end{aligned}$$

To see the second equality, write

$$\begin{aligned}
&\mathcal{H}\left(N_2X \middle| N_0X \oplus N_1\widetilde{W}, \boldsymbol{T}, \boldsymbol{N}\right) \\
&\quad = \mathsf{P}(N_0 = N_1 = N_2 = 1)\,\mathcal{H}\left(X \middle| X \oplus \widetilde{W}, \boldsymbol{T}\right) + \mathsf{P}(N_0 = 0, N_2 = 1)\,\mathcal{H}(X|\boldsymbol{T}) \\
&\quad = \mathsf{P}(N_0 = N_2 = 1)\,\mathsf{P}(N_1 = 1)\,\mathcal{H}(X|\boldsymbol{T}) + \mathsf{P}(N_0 = 1, N_2 = 0)\,\mathcal{H}(X|\boldsymbol{T}) .
\end{aligned}$$

Since $N_0$ and $N_2$ are aligned, $\mathsf{P}(N_0 = N_2 = 1) = \min(\bar{\epsilon}_0, \bar{\epsilon}_2)$ and $\mathsf{P}(N_0 = 1, N_2 = 0) = (\bar{\epsilon}_2 - \bar{\epsilon}_0)^+$. Note also that $\bar{\epsilon}_1 \min(\bar{\epsilon}_2, \bar{\epsilon}_0) + (\bar{\epsilon}_2 - \bar{\epsilon}_0)^+ = \bar{\epsilon}_1\bar{\epsilon}_2 + \epsilon_1(\bar{\epsilon}_2 - \bar{\epsilon}_0)^+ = \beta_1$ with assumption of $\bar{\epsilon}_2 \geq \bar{\epsilon}_0$. Hence the proof of Lemma 1. ■

Now back to the proof of the converse result. For each $i$, we apply Lemma 1 to (19) with

$$\boldsymbol{T}_i = \left( (N_2X)_1^{i-1}, (N_0X)_{i+1}^n, (\widetilde{Y})_1^{i-1}, (\widetilde{Y})_{i+1}^n, (\boldsymbol{N})_1^{i-1}, (\boldsymbol{N})_{i+1}^n \right)$$

which is independent of $(\boldsymbol{N})_i$. In particular,

$$\begin{aligned}
&\mathcal{H}\left( (N_0X)_i \middle| (N_2X)_1^{i-1}, (N_0X)_{i+1}^n, (\widetilde{Y})_1^n, (\boldsymbol{N})_1^n \right) \\
&= \mathcal{H}\left( (N_0X)_i \middle| (\widetilde{Y})_i, \boldsymbol{T}_i, (\boldsymbol{N})_i \right) \\
&= \frac{\alpha_1}{\beta_1} \mathcal{H}\left( (N_2X)_i \middle| (\widetilde{Y})_i, \boldsymbol{T}_i, (\boldsymbol{N})_i \right) \\
&= \frac{\alpha_1}{\beta_1} \mathcal{H}\left( (N_2X)_i \middle| (N_2X)_1^{i-1}, (N_0X)_{i+1}^n, (\widetilde{Y})_1^n, (\boldsymbol{N})_1^n \right). \qquad (20)
\end{aligned}$$

By (19) and (20),

$$\begin{aligned}
&\mathcal{H}\left( (N_2X)_1^n \middle| (\widetilde{Y})_1^n, (\boldsymbol{N})_1^n \right) - \mathcal{H}\left( (N_0X)_1^n \middle| (\widetilde{Y})_1^n, (\boldsymbol{N})_1^n \right) \\
&= \sum_{i=1}^n \left( 1 - \frac{\alpha_1}{\beta_1} \right) \mathcal{H}\left( (N_2X)_i \middle| (\widetilde{Y})_i, \boldsymbol{T}_i, (\boldsymbol{N})_i \right) \\
&\leq \sum_{i=1}^n \left( 1 - \frac{\alpha_1}{\beta_1} \right) \mathcal{H}\left( (N_2X)_i \middle| (N_2X)_1^{i-1}, (\widetilde{Y})_1^n, (\boldsymbol{N})_1^n \right) \qquad (21) \\
&= \left( 1 - \frac{\alpha_1}{\beta_1} \right) \mathcal{H}\left( (N_2X)_1^n \middle| (\widetilde{Y})_1^n, (\boldsymbol{N})_1^n \right) \qquad (22)
\end{aligned}$$

where (21) is due to the facts that conditioning reduces entropy and $\beta_1 \geq \alpha_1$ when $\bar{\epsilon}_0 \leq \bar{\epsilon}_2$, and (22) is by the chain rule. We rewrite (22) as

$$\mathcal{H}\left( (N_0X)_1^n \middle| (\widetilde{Y})_1^n, (\boldsymbol{N})_1^n \right) \geq \frac{\alpha_1}{\beta_1} \mathcal{H}\left( (N_2X)_1^n \middle| (\widetilde{Y})_1^n, (\boldsymbol{N})_1^n \right). \qquad (23)$$

Comparing (17), (18), and (23), we have

$$nR_1 + \frac{\alpha_1}{\beta_1} nR_2 - n\delta_n - \frac{\alpha_1}{\beta_1} n\delta_n \leq \bar{\epsilon}_1 + \frac{\alpha_1}{\beta_1} \bar{\epsilon}_0 \epsilon_1.$$

Sending $n \to \infty$ yields (10).

## V. Proof for the General Layered Erasure Model

In this section, we prove Theorem 2 in full generality, using insights obtained in the proof for the single-layer model. We begin with the converse.

### A. The Converse Part of Theorem 2

Because the capacity region depends only on the marginal distributions of the received signals, we can assume arbitrary joint distribution of the fading coefficients $N_0$ and $N_2$ as long as the marginals remain the same. Throughout the proof of the converse of Theorem 2 (Section V-A), it is assume that $\{N_1[m]\}$ is independent of $\{N_0[m], N_2[m]\}$, and $\{N_0[m]\}$ and $\{N_2[m]\}$ are aligned as described in the following. Let $F_N(n) = \mathsf{P}\left(N \leq n\right)$ denote the cumulative distribution function of an arbitrary random variable $N$, and define its inverse as $F_N^{-1}(t) = \inf\{u : F_N(u) \geq t\}$. Let $\Lambda$ be a uniform random variable on $[0,1]$, then $F_N^{-1}(\Lambda)$ is identically distributed as $N$. Let

$$N_0[m] = F_{N_0}^{-1}(\Lambda[m]) \ , \quad \text{and} \quad N_2[m] = F_{N_2}^{-1}(\Lambda[m]) \ , \quad m = 1, 2, \dots \tag{24}$$

where $\{\Lambda[m]\}$ are i.i.d. and uniformly distributed on $[0,1]$. Basically, once aligned, a larger realization of $N_0$ implies a larger realization of $N_2$, and vice verse. Note that, unlike in single-layer case of $q = 1$, there is no guarantee that one of the fading coefficients dominates the other (*e.g.*, $N_0 \equiv 1$ but $N_2$ can take values of 0 and 2). Therefore, the layered interference channel cannot always be categorized as a strong or weak interference channel.

It is important to note that $\alpha_l$ and $\beta_l$, $l = 1, \dots, q$, as well as the region $\mathcal{C}$, remain unchanged after the alignment. In particular, only the marginals distributions of $N_0$ and $N_2$ are used in the definition of $\beta_l$.

Let the elements of $(\widetilde{\boldsymbol{W}})_1^n$, *i.e.*, $\widetilde{W}_i[j]$, $i = 1, \dots, q$ and $j = 1, \dots, n$, be i.i.d. Ber$\,(1/2)$ random variables. Also, $(\widetilde{\boldsymbol{W}})_1^n$ is independent of all channel inputs and states. The following generalization of Lemma 1 holds.

*Lemma 2:* Suppose $N_0$ and $N_2$ are aligned according to (24). Let $\boldsymbol{T}$ be a collection of random variables independent of $\boldsymbol{N} = (N_0, N_1, N_2)$. Let $\boldsymbol{X}$ be an arbitrary random vector in $\mathbb{F}_2^q$ independent of $\widetilde{\boldsymbol{W}}$. Then

$$\mathcal{H}\left(\boldsymbol{X}_1^{N_0} \middle| \boldsymbol{X}_1^{N_0} \oplus \widetilde{\boldsymbol{W}}_1^{N_1}, \boldsymbol{T}, \boldsymbol{N}\right) = \sum_{l=1}^{q} \alpha_l \mathcal{H}\left(X_l \middle| \boldsymbol{X}_1^{l-1}, \boldsymbol{T}\right) \tag{25}$$

$$\mathcal{H}\left(\boldsymbol{X}_1^{N_2} \middle| \boldsymbol{X}_1^{N_0} \oplus \widetilde{\boldsymbol{W}}_1^{N_1}, \boldsymbol{T}, \boldsymbol{N}\right) = \sum_{l=1}^{q} \beta_l \mathcal{H}\left(X_l \middle| \boldsymbol{X}_1^{l-1}, \boldsymbol{T}\right) \tag{26}$$

where $\alpha_l$ and $\beta_l$ are given by (7) and (8), respectively.

The proof of lemma 2 is based on direct computation. For details, see Appendix I. It is interesting to note that the expression of $\beta_l$ given in (8) is independent of the correlation between $N_0$ and $N_2$, although the derivation depends on the alignment between them. Therefore, the expression of capacity region (6) does not depend on the artificial alignment either.

An interpretation for Lemma 2 is as follows. Suppose that we can observe $\boldsymbol{X}$ through three channels: $\boldsymbol{X}_1^{N_0}$, $\boldsymbol{X}_1^{N_2}$, and $\boldsymbol{X}_1^{N_0} \oplus \widetilde{\boldsymbol{W}}_1^{N_1}$. By (7), $\alpha_l$ is the probability that layer $l$ of $\boldsymbol{X}$ can be seen in $\boldsymbol{X}_1^{N_0}$ but not through the channel $\boldsymbol{X}_1^{N_0} \oplus \widetilde{\boldsymbol{W}}_1^{N_1}$. Whenever this event happens, the amount of entropy $\mathcal{H}\left(X_l \middle| \boldsymbol{X}_1^{l-1}, \boldsymbol{T}\right)$ is accumulated (via chain rule). Hence, (25) follows. Similar interpretation can be obtain for (26) by noting that $\beta_l$ is the probability that layer $l$ of $\boldsymbol{X}$ can be seen in $\boldsymbol{X}_1^{N_2}$ but not through the channel $\boldsymbol{X}_1^{N_0} \oplus \widetilde{\boldsymbol{W}}_1^{N_1}$ under the assumption of alignment between $N_0$ and $N_2$.

Equipped with Lemma 2, we next establish the constraints on the capacity region in Theorem 2. The first constraint in (6) is trivial by cut-set bound. The second constraint for $R_1 + \omega R_2$ is proved in the following.

For notational convenience, let $\widetilde{\boldsymbol{Y}} = \widetilde{\boldsymbol{W}}_1^{N_1} \oplus \boldsymbol{X}_1^{N_0}$. By Fano's inequality,

$$
\begin{aligned}
nR_1 - n\delta_n &\le \mathcal{I}\left((\boldsymbol{Y})_1^n; (\boldsymbol{W})_1^n | (\boldsymbol{N})_1^n\right) \\
&= \mathcal{H}\left((\boldsymbol{W}_1^{N_1} \oplus \boldsymbol{X}_1^{N_0})_1^n \middle| (\boldsymbol{N})_1^n\right) - \mathcal{H}\left((\boldsymbol{X}_1^{N_0})_1^n \middle| (\boldsymbol{N})_1^n\right) \\
&\le \mathcal{H}\left(\left(\widetilde{\boldsymbol{W}}_1^{N_1} \oplus \boldsymbol{W}_1^{N_1} \oplus \boldsymbol{X}_1^{N_0}\right)_1^n \middle| (\boldsymbol{N})_1^n\right) - \mathcal{H}\left((\boldsymbol{X}_1^{N_0})_1^n \middle| (\boldsymbol{N})_1^n\right) \\
&= \mathcal{H}\left((\widetilde{\boldsymbol{W}}_1^{N_1} \oplus \boldsymbol{X}_1^{N_0})_1^n \middle| (\boldsymbol{N})_1^n\right) - \mathcal{H}\left((\boldsymbol{X}_1^{N_0})_1^n \middle| (\boldsymbol{N})_1^n\right) && (27) \\
&= \mathcal{I}\left((\widetilde{\boldsymbol{Y}})_1^n; (\widetilde{\boldsymbol{W}}_1^{N_1})_1^n \middle| (\boldsymbol{N})_1^n\right) \\
&= \mathcal{H}\left((\widetilde{\boldsymbol{W}}_1^{N_1})_1^n \middle| (\boldsymbol{N})_1^n\right) - \mathcal{H}\left((\widetilde{\boldsymbol{W}}_1^{N_1})_1^n \middle| (\widetilde{\boldsymbol{Y}})_1^n, (\boldsymbol{N})_1^n\right) \\
&= n\mathbb{E}N_1 - \mathcal{H}\left((\boldsymbol{X}_1^{N_0})_1^n \middle| (\widetilde{\boldsymbol{Y}})_1^n, (\boldsymbol{N})_1^n\right) && (28)
\end{aligned}
$$

where (27) is because $\boldsymbol{W}_1^{N_1} \oplus \widetilde{\boldsymbol{W}}_1^{N_1}$ is identically distributed as $\widetilde{\boldsymbol{W}}_1^{N_1}$.

Also by Fano's inequality, and by the chain rule of mutual information,

$$
\begin{aligned}
nR_2 - n\delta_n &\le \mathcal{I}\left((\boldsymbol{X}_1^{N_2})_1^n; (\boldsymbol{X})_1^n \middle| (\boldsymbol{N})_1^n\right) \\
&\le \mathcal{I}\left((\boldsymbol{X}_1^{N_2})_1^n, (\widetilde{\boldsymbol{Y}})_1^n; (\boldsymbol{X})_1^n \middle| (\boldsymbol{N})_1^n\right) \\
&= \mathcal{I}\left((\widetilde{\boldsymbol{Y}})_1^n; (\boldsymbol{X})_1^n \middle| (\boldsymbol{N})_1^n\right) + \mathcal{I}\left((\boldsymbol{X}_1^{N_2})_1^n; (\boldsymbol{X})_1^n \middle| (\widetilde{\boldsymbol{Y}})_1^n, (\boldsymbol{N})_1^n\right) \\
&= \mathcal{I}\left((\widetilde{\boldsymbol{Y}})_1^n; (\boldsymbol{X})_1^n \middle| (\boldsymbol{N})_1^n\right) + \mathcal{H}\left((\boldsymbol{X}_1^{N_2})_1^n \middle| (\widetilde{\boldsymbol{Y}})_1^n, (\boldsymbol{N})_1^n\right) \quad . && (29)
\end{aligned}
$$

By the property of memoryless channels, the first term on the right hand side (RHS) of (29) can be upper bounded:

$$\begin{aligned}\mathcal{I}\left((\widetilde{\boldsymbol{Y}})_1^n;(\boldsymbol{X})_1^n\middle|(\boldsymbol{N})_1^n\right) &\leq \sum_{i=1}^{n}\mathcal{I}\left((\widetilde{\boldsymbol{Y}})_i;(\boldsymbol{X})_i\middle|(\boldsymbol{N})_1^n\right)\\ &\leq \sum_{i=1}^{n}\mathcal{H}\left((\widetilde{\boldsymbol{Y}})_i\middle|(\boldsymbol{N})_1^n\right)-\mathcal{H}\left((\widetilde{\boldsymbol{W}}_1^{N_1})_i\middle|(\boldsymbol{N})_1^n\right)\\ &\leq n\mathbb{E}\max(N_1,N_0)-n\mathbb{E}N_1 \end{aligned}\tag{30}$$

By (29) and (30), we have

$$nR_2-n\delta_n\leq n\mathbb{E}\left[N_0-N_1\right]^+ + \mathcal{H}\left((\boldsymbol{X}_1^{N_2})_1^n\middle|(\widetilde{\boldsymbol{Y}})_1^n,(\boldsymbol{N})_1^n\right)\ .\tag{31}$$

Note that (28) and (31) have the same interpretation as (17) and (18), respectively. By (28) and (31), we have the following weighted bound for every $\omega\in[0,1]$,

$$\begin{aligned}&nR_1+n\omega R_2-(1+\omega)n\delta_n\leq n\mathbb{E}N_1+n\omega\mathbb{E}\left[N_0-N_1\right]^+\\ &\qquad -\mathcal{H}\left((\boldsymbol{X}_1^{N_0})_1^n\middle|(\widetilde{\boldsymbol{Y}})_1^n,(\boldsymbol{N})_1^n\right)+\omega\mathcal{H}\left((\boldsymbol{X}_1^{N_2})_1^n\middle|(\widetilde{\boldsymbol{Y}})_1^n,(\boldsymbol{N})_1^n\right)\ .\end{aligned}\tag{32}$$

Similar to what was shown in the case of $q=1$, we use a "Marton-like" expansion to write the difference of the two entropies on the RHS of (32) in the special case of $\omega=1$:

$$\begin{aligned}&\mathcal{H}\left((\boldsymbol{X}_1^{N_2})_1^n\middle|(\widetilde{\boldsymbol{Y}})_1^n,(\boldsymbol{N})_1^n\right)-\mathcal{H}\left((\boldsymbol{X}_1^{N_0})_1^n\middle|(\widetilde{\boldsymbol{Y}})_1^n,(\boldsymbol{N})_1^n\right)\\ &=\sum_{i=1}^{n}\Big\{\mathcal{H}\left((\boldsymbol{X}_1^{N_2})_1^i,(\boldsymbol{X}_1^{N_0})_{i+1}^n\middle|(\widetilde{\boldsymbol{Y}})_1^n,(\boldsymbol{N})_1^n\right)\\ &\qquad -\mathcal{H}\left((\boldsymbol{X}_1^{N_2})_1^{i-1},(\boldsymbol{X}_1^{N_0})_i^n\middle|(\widetilde{\boldsymbol{Y}})_1^n,(\boldsymbol{N})_1^n\right)\Big\}\\ &=\sum_{i=1}^{n}\Big\{\mathcal{H}\left((\boldsymbol{X}_1^{N_2})_1^{i-1},(\boldsymbol{X}_1^{N_0})_{i+1}^n\middle|(\widetilde{\boldsymbol{Y}})_1^n,(\boldsymbol{N})_1^n\right)\\ &\qquad +\mathcal{H}\left((\boldsymbol{X}_1^{N_2})_i\middle|(\boldsymbol{X}_1^{N_2})_1^{i-1},(\boldsymbol{X}_1^{N_0})_{i+1}^n,(\widetilde{\boldsymbol{Y}})_1^n,(\boldsymbol{N})_1^n\right)\\ &\qquad\qquad -\mathcal{H}\left((\boldsymbol{X}_1^{N_2})_1^{i-1},(\boldsymbol{X}_1^{N_0})_{i+1}^n\middle|(\widetilde{\boldsymbol{Y}})_1^n,(\boldsymbol{N})_1^n\right)\\ &\qquad\qquad\qquad -\mathcal{H}\left((\boldsymbol{X}_1^{N_0})_i\middle|(\boldsymbol{X}_1^{N_2})_1^{i-1},(\boldsymbol{X}_1^{N_0})_{i+1}^n(\widetilde{\boldsymbol{Y}})_1^n,(\boldsymbol{N})_1^n\right)\Big\}\\ &=\sum_{i=1}^{n}\Big\{\mathcal{H}\left((\boldsymbol{X}_1^{N_2})_i\middle|(\widetilde{\boldsymbol{Y}})_i,\boldsymbol{T}_i,(\boldsymbol{N})_i\right)-\mathcal{H}\left((\boldsymbol{X}_1^{N_0})_i\middle|(\widetilde{\boldsymbol{Y}})_i,\boldsymbol{T}_i,(\boldsymbol{N})_i\right)\Big\}\end{aligned}\tag{33}$$

where

$$\boldsymbol{T}_i = \left( (\boldsymbol{X}_1^{N_2})_1^{i-1}, (\boldsymbol{X}_1^{N_0})_{i+1}^n, (\widetilde{\boldsymbol{Y}})_1^{i-1}, (\widetilde{\boldsymbol{Y}})_{i+1}^n, (\boldsymbol{N})_1^{i-1}, (\boldsymbol{N})_{i+1}^n \right) \tag{34}$$

which is independent of $(\boldsymbol{N})_i$. Note also that

$$\begin{aligned}
&\mathcal{H}\left( (\boldsymbol{X}_1^{N_2})_1^n \middle| (\widetilde{\boldsymbol{Y}})_1^n, (\boldsymbol{N})_1^n \right) \\
&\quad = \sum_{i=1}^n \mathcal{H}\left( (\boldsymbol{X}_1^{N_2})_i \middle| (\boldsymbol{X}_1^{N_2})_1^{i-1}, (\widetilde{\boldsymbol{Y}})_1^n, (\boldsymbol{N})_1^n \right) \\
&\quad \geq \sum_{i=1}^n \mathcal{H}\left( (\boldsymbol{X}_1^{N_2})_i \middle| (\widetilde{\boldsymbol{Y}})_i, \boldsymbol{T}_i, (\boldsymbol{N})_i \right) .
\end{aligned} \tag{35}$$

Therefore, by (33) and (35), the difference on the RHS of (32) for any $\omega \in [0,1]$ can be written as

$$\begin{aligned}
&\omega\mathcal{H}\left( (\boldsymbol{X}_1^{N_2})_1^n \middle| (\widetilde{\boldsymbol{Y}})_1^n, (\boldsymbol{N})_1^n \right) - \mathcal{H}\left( (\boldsymbol{X}_1^{N_0})_1^n \middle| (\widetilde{\boldsymbol{Y}})_1^n, (\boldsymbol{N})_1^n \right) \\
&= \mathcal{H}\left( (\boldsymbol{X}_1^{N_2})_1^n \middle| (\widetilde{\boldsymbol{Y}})_1^n, (\boldsymbol{N})_1^n \right) - \mathcal{H}\left( (\boldsymbol{X}_1^{N_0})_1^n \middle| (\widetilde{\boldsymbol{Y}})_1^n, (\boldsymbol{N})_1^n \right) \\
&\qquad - (1-\omega)\mathcal{H}\left( (\boldsymbol{X}_1^{N_2})_1^n \middle| (\widetilde{\boldsymbol{Y}})_1^n, (\boldsymbol{N})_1^n \right) \\
&\leq \sum_{i=1}^n \left\{ \mathcal{H}\left( (\boldsymbol{X}_1^{N_2})_i \middle| (\widetilde{\boldsymbol{Y}})_i, \boldsymbol{T}_i, (\boldsymbol{N})_i \right) - \mathcal{H}\left( (\boldsymbol{X}_1^{N_0})_i \middle| (\widetilde{\boldsymbol{Y}})_i, \boldsymbol{T}_i, (\boldsymbol{N})_i \right) \right\} \\
&\qquad - (1-\omega) \sum_{i=1}^n \mathcal{H}\left( (\boldsymbol{X}_1^{N_2})_i \middle| (\widetilde{\boldsymbol{Y}})_i, \boldsymbol{T}_i, (\boldsymbol{N})_i \right) \\
&= \sum_{i=1}^n \left\{ \omega\mathcal{H}\left( (\boldsymbol{X}_1^{N_2})_i \middle| (\widetilde{\boldsymbol{Y}})_i, \boldsymbol{T}_i, (\boldsymbol{N})_i \right) - \mathcal{H}\left( (\boldsymbol{X}_1^{N_0})_i \middle| (\widetilde{\boldsymbol{Y}})_i, \boldsymbol{T}_i, (\boldsymbol{N})_i \right) \right\} .
\end{aligned}$$

Applying Lemma 2 to both entropy terms with $\boldsymbol{T}_i$ defined in (34) yields

$$\begin{aligned}
&\omega\mathcal{H}\left( (\boldsymbol{X}_1^{N_2})_1^n \middle| (\widetilde{\boldsymbol{Y}})_1^n, (\boldsymbol{N})_1^n \right) - \mathcal{H}\left( (\boldsymbol{X}_1^{N_0})_1^n \middle| (\widetilde{\boldsymbol{Y}})_1^n, (\boldsymbol{N})_1^n \right) \\
&\qquad \leq \sum_{i=1}^n \sum_{l=1}^q (\omega\beta_l - \alpha_l) \mathcal{H}\left( (X_l)_i \middle| (\boldsymbol{X}_1^{l-1})_i, \boldsymbol{T}_i \right) \\
&\qquad \leq n \sum_{l=1}^q (\omega\beta_l - \alpha_l)^+ .
\end{aligned} \tag{36}$$

Therefore, substituting (36) into (32) and noting that $\delta_n \to 0$ as $n \to \infty$, we have established the converse part in Theorem 2.

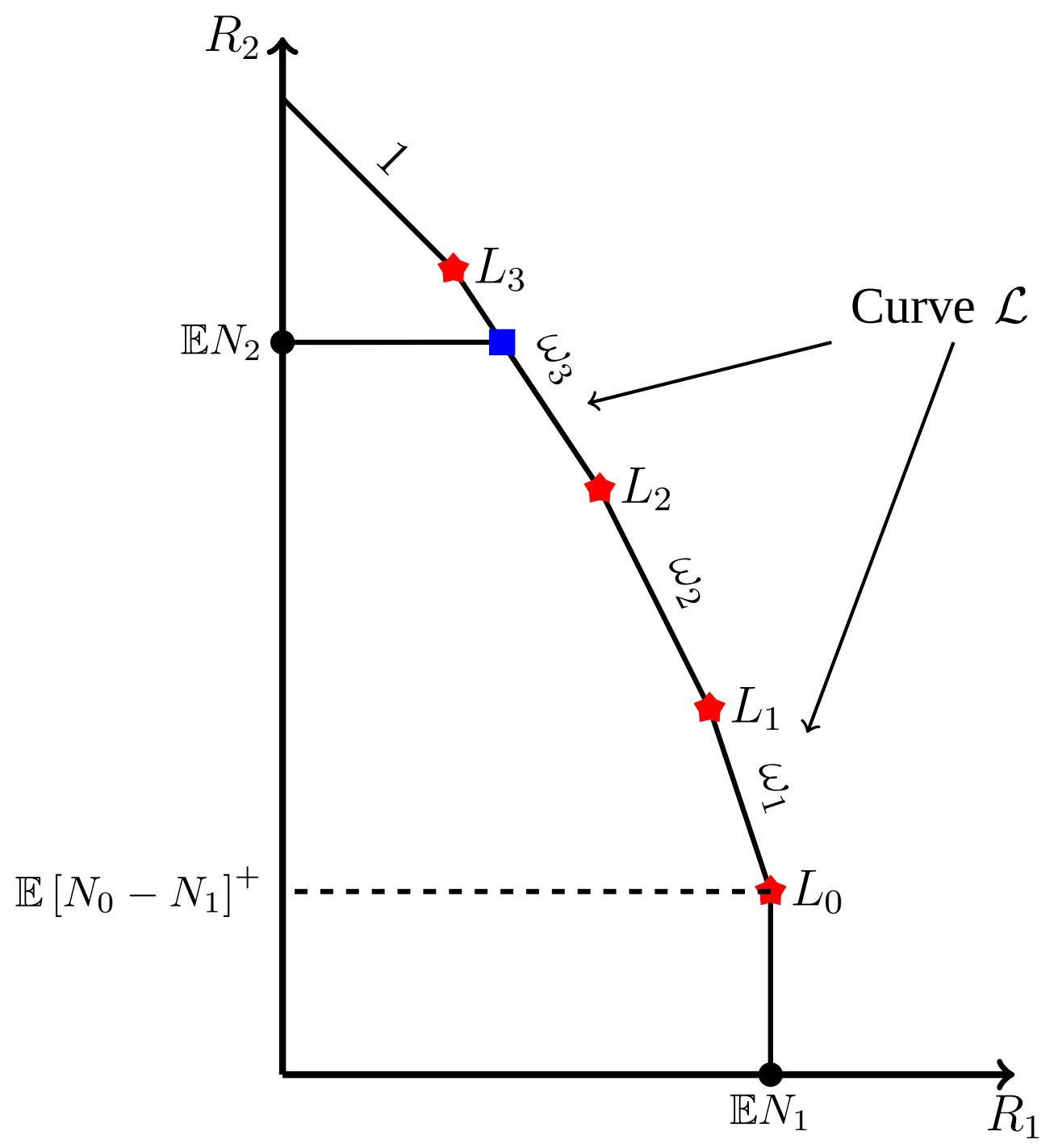

Fig. 3. An illustration of the capacity region for general layered erasure channel. The region is generally enclosed by the axes, line $R_2 = \mathbb{E}N_2$, and a piece-wisely linear curve $\mathcal{L}$. The top maximum-sum-rate point is marked by square, whose position variates for different cases.

### B. *The Achievability Part of Theorem 2*

Let us first investigate the geometry of the region $\mathcal{C}$ given by (6). Assume $\mathbb{E}N_1, \mathbb{E}N_2 \neq 0$; otherwise, the capacity region is trivial. The region bounded by the second constraint in (6) can be viewed as $\cap_{\omega\in[0,1]} H(\omega) \cap [0, +\infty)^2$ where

$$H(\omega) = \left\{ (R_1, R_2) \middle| R_1 + \omega R_2 \leq \mathbb{E}N_1 + \omega\mathbb{E}[N_0 - N_1]^+ + \sum_{l\in\mathcal{B}(\omega)} (\omega\beta_l - \alpha_l)^+ \right\}$$

and $\mathcal{B}(\omega) = \{l \in \{1, \cdots, q\} \,|\, \omega\beta_l \geq \alpha_l\}$. Let us order $\{\alpha_l/\beta_l \,|\, l = 1, \ldots, q\}$ as $\omega_1 \leq \cdots \leq \omega_b < 1 \leq \omega_{b+1} \cdots \leq \omega_q$, and let the corresponding permutation be referred to as $\tau$ so that $\omega_i = \alpha_{\tau(i)}/\beta_{\tau(i)}$, $i = 1, \ldots, q$. In addition, let $\omega_0 = 0$. It turns out that except for the $b + 2$ constraints $H(\omega_k)$, $i = 0, \ldots, b$, and $H(1)$, all constraints $H(\omega)$ with other $\omega$ are redundant:

*Proposition 2:*

$$\bigcap_{\omega\in[0,1]} H(\omega) = H(\omega_0) \cap \cdots \cap H(\omega_b) \cap H(1) \tag{37}$$

*Proof:* For every $\omega \in [0,1]$, $H(\omega)$ is a half plane to the left side of a straight line. For $i = 0, \dots, b-1$, the boundary of $H(\omega_i)$ and the boundary of $H(\omega_{i+1})$ intersect at the following point:

$$\left( \mathbb{E}N_1 - \sum_{l \in \mathcal{B}(\omega_i)} \alpha_l \ , \quad \mathbb{E}[N_0 - N_1]^+ + \sum_{l \in \mathcal{B}(\omega_i)} \beta_l \right) \tag{38}$$

which is denoted by $L_i$ from now on. In particular, this is because for layer $l$ in $\mathcal{B}(\omega_{i+1})$ but not in $\mathcal{B}(\omega_i)$, we have $\omega_{i+1}\beta_l - \alpha_l = 0$. In addition, let $L_b$ be the intersection point of the boundary of $H(\omega_b)$ and the boundary of $H(1)$, whose coordinate is also given by (38) by changing the subscript $i$ to $b$. Define intervals $\Omega_i = (\omega_i, \omega_{i+1})$ for $i = 0, \dots, b-1$, and $\Omega_b = (\omega_b, 1)$. For every $i = 0, \dots, b$, it is not difficult to see that $\mathcal{B}(\omega) = \mathcal{B}(\omega_i)$ for all $\omega \in \Omega_i$. Furthermore, the boundary of $H(\omega)$ also passes point $L_i$. Thus we see that the constraint $H(\omega)$ is redundant to $H(\omega_i)$ and $H(\omega_{i+1})$ (or $H(1)$ for $i = b$). Hence the proof of the proposition. ■

In this one-sided interference channel problem, the parameter $\omega$ can be interpreted as a preference between rate loss of user 1 and rate gain of user 2 in view of (28) and (31). This is in contrast to the layered erasure broadcast channel problem studied in [17], where the role of the weighting parameter in a weighted sum-rate characterization of the capacity region is interpreted as a preference between the two users.

In general, the upper bound of the second constraint in (6), henceforth referred to by the boundary $\mathcal{L}$, is piece-wisely straight, and its corner points are $L_0, \dots, L_b$. Note the point $L_{i+1}$ is always on the upper left side of $L_i$. In case of degeneracy, the two points coincide. The region $\mathcal{C}$ is the region enclosed by $\mathcal{L}$, the line $R_2 = \mathbb{E}N_2$, and the two axes. The line $R_2 = \mathbb{E}N_2$ can intersect with the curve $\mathcal{L}$ at various positions, and we call the intersection the *top maximum-sum-rate point* for obvious reason. To prove the achievability of the region, it suffices to show that the top maximum-sum-rate point and all corner points $L_i$ below it are achievable.

*1) The Corner Points $L_0, \dots, L_b$:* For each $i \in \{0, \dots, b\}$, consider the achievability of the corner point $L_i$ which is below the top maximum-sum-rate point. Let user 1 generate a random codebook of rate

$$R_1 = \mathbb{E}N_1 - \sum_{l \in \mathcal{B}(\omega_i)} \alpha_l$$

and let user 2 generate two codebooks: one is for private message at rate

$$R_{2p} = \sum_{l \in \mathcal{B}(\omega_i)} \mathsf{P}\left(N_2 \geq l\right) \tag{39}$$

the other for common message at rate

$$R_{2c} = \sum_{l \in \mathcal{U}(\omega_i)} \mathsf{P}\left(N_0 - N_1 \geq l\right) \tag{40}$$

where $\mathcal{U}(\omega) = \{1, \ldots, q\} \backslash \mathcal{B}(\omega)$ for every $\omega \in [0,1]$. For notational convenience, for a random vector $\boldsymbol{X} \in \mathbb{F}_2^q$ and a subset of $\{1, \ldots, q\}$, $\mathcal{A}$, we denote $\boldsymbol{X}_{\mathcal{A}}$ as a $q$-dimensional vector whose $l$th element is $X_l$ if $l \in \mathcal{A}$ and equals to $0$, otherwise. All codebooks consist of i.i.d. $\mathsf{Ber}\,(1/2)$ entries. The codeword of user 1 is transmitted as $\boldsymbol{W}$. The codeword for the private message of user 2 is transmitted as $\boldsymbol{X}_{\mathcal{B}(\omega_i)}$ using the layers in $\mathcal{B}(\omega_i)$, whereas the codeword for the common message is transmitted as $\boldsymbol{X}_{\mathcal{U}(\omega_i)}$ using the remaining layers.

Note that, by (7), we can also write $\alpha_l = \mathbb{E}\big[\mathsf{P}\left(N_0 \geq l\right) - \mathsf{P}\left(N_0 - N_1 \geq l | N_1\right)\big]$. Comparing with (8), we find that for every $l \in \mathcal{B}(\omega_i)$, $\beta_l \geq \alpha_l$ since $\omega_i \leq 1$, so that $\mathsf{P}\left(N_2 \geq l\right) \geq \mathsf{P}\left(N_0 \geq l\right) \geq \mathsf{P}\left(N_0 - N_1 \geq l | N_1\right)$, which implies that $\beta_l = \mathsf{P}\left(N_2 \geq l\right) - \mathsf{P}\left(N_0 - N_1 \geq l\right)$. Therefore, by (39) and (40),

$$\begin{aligned}
R_{2c} + R_{2p} &= \sum_{l \in \mathcal{B}(\omega_i)} \mathsf{P}\left(N_2 \geq l\right) + \sum_{l \in \mathcal{U}(\omega_i)} \mathsf{P}\left(N_0 - N_1 \geq l\right) \\
&= \sum_{l=1}^{q} \mathsf{P}\left(N_0 - N_1 \geq l\right) + \sum_{l \in \mathcal{B}(\omega_i)} \mathsf{P}\left(N_2 \geq l\right) - \sum_{l \in \mathcal{B}(\omega_i)} \mathsf{P}\left(N_0 - N_1 \geq l\right) \\
&= \mathbb{E}[N_0 - N_1]^+ + \sum_{l \in \mathcal{B}(\omega_i)} \beta_l \;.
\end{aligned}$$

Thus by (38), the codebooks carry exactly the rate pair at $L_i$. Therefore, to show achievability of point $L_i$ is equivalent to show that the rate triple $(R_1, R_{2p}, R_{2c})$ is achievable. Indeed, receiver 1 can decode the common message, because

$$\begin{aligned}
\mathcal{I}\left(\boldsymbol{W}_1^{N_1} \oplus \boldsymbol{X}_1^{N_0}; \boldsymbol{X}_{\mathcal{U}(\omega_i)} | N_0, N_1\right) &= \sum_{l \in \mathcal{U}(\omega_i)} \mathcal{I}\left(\boldsymbol{W}_1^{N_1} \oplus \boldsymbol{X}_1^{N_0}; X_l | N_0, N_1\right) \\
&= \sum_{l \in \mathcal{U}(\omega_i)} \mathsf{P}\left(N_0 - N_1 \geq l\right) \\
&= R_{2c}\,.
\end{aligned}$$

Intuitively, for given $N_0 = n_0$ and $N_1 = n_1$, the signal $X_l$ at level $l$ in $\mathcal{U}(\omega_i)$ contributes to the mutual information if and only if $n_0 - n_1 \geq l$. After canceling the interference caused by the common message, receiver 1 can decode its own message, because

$$\begin{aligned}
&\mathcal{I}\left(\boldsymbol{W}_1^{N_1} \oplus \boldsymbol{X}_1^{N_0}; \boldsymbol{W} \middle| \boldsymbol{X}_{\mathcal{U}(\omega_i)}, N_1, N_0\right) \\
&\quad = \mathcal{H}\left(\boldsymbol{W}_1^{N_1} \middle| N_1, N_0\right) - \mathcal{H}\left(\boldsymbol{W}_1^{N_1} \middle| \boldsymbol{W}_1^{N_1} \oplus \boldsymbol{X}_1^{N_0}, \boldsymbol{X}_{\mathcal{U}(\omega_i)}, N_1, N_0\right) \\
&\quad = \mathbb{E}N_1 - \mathcal{H}\left(\boldsymbol{X}_1^{N_0} \middle| \boldsymbol{W}_1^{N_1} \oplus \boldsymbol{X}_1^{N_0}, \boldsymbol{X}_{\mathcal{U}(\omega_i)}, N_1, N_0\right) \\
&\quad = \mathbb{E}N_1 - \sum_{l \in \mathcal{B}(\omega_i)} \alpha_l \qquad (41) \\
&\quad = R_1
\end{aligned}$$

where (41) can be regarded as a consequence of Lemma 2. Receiver 2 can decode its private message because

$$\begin{aligned}
\mathcal{I}\left(\boldsymbol{X}_1^{N_2}; \boldsymbol{X}_{\mathcal{B}(\omega_i)} \middle| N_2\right) &= \sum_{l \in \mathcal{B}(\omega_i)} \mathcal{I}\left(\boldsymbol{X}_1^{N_2}; X_l \middle| N_2\right) \\
&= \sum_{l \in \mathcal{B}(\omega_i)} \mathsf{P}\left(N_2 \geq l\right) \\
&= R_{2p}\,.
\end{aligned}$$

Receiver 2 can also decode the common message:

$$\begin{aligned}
\mathcal{I}\left(\boldsymbol{X}_1^{N_2}; \boldsymbol{X}_{\mathcal{U}(\omega_i)} | N_2\right) &= \sum_{l \notin \mathcal{B}(\omega_i)} \mathsf{P}\left(N_2 \geq l\right) \\
&= \mathbb{E}N_2 - R_{2p} \\
&\geq R_{2c}
\end{aligned}$$

where the inequality is because $L_i$ is below the top maximum-sum-rate point, i.e., $R_{2p} + R_{2c} \leq \mathbb{E}N_2$.

*2) The Top Maximum-sum-rate Point:* We establish the achievability of the top maximum-sum-rate point in all three possible cases depending its position.

Case 1: The top maximum-sum-rate point is below $L_0$. In this case, $\mathbb{E}[N_0 - N_1]^+ \geq \mathbb{E}N_2$, so that the region (6) becomes rectangular. It suffices to show that $(\mathbb{E}N_1, \mathbb{E}N_2)$ is achievable. Let user 1 and user 2 each generate a random codebook with i.i.d. Ber$(1/2)$ entries, with rate $\mathbb{E}N_1$ and $\mathbb{E}N_2$, respectively. Then receiver 1 can decode the message of user 2, because

$$\begin{aligned}\mathcal{I}\left(\boldsymbol{W}_1^{N_1} \oplus \boldsymbol{X}_1^{N_0}; \boldsymbol{X}|N_1,N_0\right) &= \sum_{l=1}^{q} \mathcal{I}\left(\boldsymbol{W}_1^{N_1} \oplus \boldsymbol{X}_1^{N_0}; X_l|N_1,N_0\right) \\ &= \sum_{l=1}^{q} \mathsf{P}\left(N_0 - N_1 \geq l\right) \\ &= \mathbb{E}[N_0 - N_1]^+ \\ &\geq \mathbb{E}N_2\,.\end{aligned}$$

After canceling the interference, receiver 1 can decode its own message, because

$$\begin{aligned}\mathcal{I}\left(\boldsymbol{W}_1^{N_1} \oplus \boldsymbol{X}_1^{N_0}; \boldsymbol{W}|\boldsymbol{X},N_1,N_0\right) &= \mathcal{I}\left(\boldsymbol{W}_1^{N_1}; \boldsymbol{W}|N_1,N_0\right) \\ &= \mathbb{E}N_1\,.\end{aligned}$$

Also, receiver 2 can decode its own message because

$$\begin{aligned}\mathcal{I}\left(\boldsymbol{Z}; \boldsymbol{X}|N_2\right) &= \mathcal{H}\left(\boldsymbol{X}_1^{N_2}|N_2\right) \\ &= \mathbb{E}N_2.\end{aligned}$$

Case 2: The top maximum-sum-rate point is between $L_0$ and $L_b$, *i.e.*, the intersection of line $R_2 = \mathbb{E}N_2$ and the curve $\mathcal{L}$ is on the boundary of $H(\omega_k)$ for some $k \in \{1,\ldots,b\}$. The basic idea of the coding scheme is to transmit a private message using layers $\tau(1),\ldots,\tau(k-1)$ and part of layer $\tau(k)$, and to transmit a common message using layers $\tau(k+1),\ldots,\tau(q)$ and the remaining part of layer $\tau(k)$.

Let user 1 generate a random codebook with i.i.d. Ber$(1/2)$ entries. Define set $\mathcal{B}' = \{\tau(1),\ldots,\tau(k-1)\}$ and set $\mathcal{U}' = \{\tau(k+1),\ldots,\tau(q)\}$. Let user 2 encode its common message onto $(\boldsymbol{X}_{\mathcal{U}'}, U)$, and encode its private message onto $(\boldsymbol{X}_{\mathcal{B}'}, V)$, where $U$ and $V$ are two binary signals. The transmitted signal $\boldsymbol{X}$ then consists of $\boldsymbol{X}_{\mathcal{U}'}$, $\boldsymbol{X}_{\mathcal{B}'}$ and $X_{\tau(k)} = \max(U,V)$. The codebook for the common message consists of random independent entries, where the elements of $\boldsymbol{X}_{\mathcal{U}'}$ are Ber$(1/2)$ and $U \sim$ Ber$(\delta/2)$ for some $\delta \in [0,1]$. The codebook for the private message is generated similarly, but with $V \sim$ Ber$(1 - 1/(2-\delta))$. We note that

$$\begin{aligned}\mathcal{H}\left(X_{\tau(k)}|U\right) &= \mathsf{P}\left(U=0\right)\mathcal{H}\left(V\right) \\ &= \frac{2-\delta}{2}\mathcal{H}\left(\frac{1}{2-\delta}\right)\ .\end{aligned} \tag{42}$$

For fixed $\delta$, the common message can be decoded at receiver 1 as long as its rate does not exceed

$$\begin{aligned}
R_{2c}(\delta) &= \mathcal{I}\left(\boldsymbol{W}_1^{N_1} \oplus \boldsymbol{X}_1^{N_0}; \boldsymbol{X}_{\mathcal{U}'}, U \middle| N_1, N_0\right) \\
&= \mathbb{E}\sum_{l=1}^{q} 1_{(l\le N_0-N_1)}\mathcal{I}\left(X_l; \boldsymbol{X}_{\mathcal{U}'}, U\right) \\
&= \mathbb{E}\sum_{l=1}^{q} 1_{(l\le N_0-N_1)}(1-\mathcal{H}\left(X_l|\boldsymbol{X}_{\mathcal{U}'}, U\right)) \\
&= \mathbb{E}[N_0-N_1]^+ - \mathbb{E}\left\{1_{(\tau(k)\le N_0-N_1)}\mathcal{H}\left(X_{\tau(k)}|U\right)\right\} - \mathbb{E}\sum_{l\in\mathcal{B}'} 1_{(l\le N_0-N_1)} \\
&= \mathbb{E}[N_0-N_1]^+ - \mathsf{P}\left(N_0-N_1\ge\tau(k)\right)\frac{2-\delta}{2}\mathcal{H}\left(\frac{1}{2-\delta}\right) - \sum_{l\in\mathcal{B}'}\mathsf{P}\left(N_0-N_1\ge l\right)
\end{aligned}$$

where (42) is used to reach the last equality. Once the interference caused by the common message is removed, receiver 1 can decode its own message at rate

$$\begin{aligned}
R_1(\delta) &= \mathcal{I}\left(\boldsymbol{W}_1^{N_1} \oplus \boldsymbol{X}_1^{N_0}; \boldsymbol{W}_1^{N_1} \middle| \boldsymbol{X}_{\mathcal{U}}, U, N_1, N_0\right) \\
&= \mathcal{H}\left(\boldsymbol{W}_1^{N_1}|N_1\right) - \mathcal{H}\left(\boldsymbol{W}_1^{N_1} \middle| \boldsymbol{W}_1^{N_1}\oplus\boldsymbol{X}_1^{N_0}, \boldsymbol{X}_{\mathcal{U}}, U, N_1, N_0\right) \\
&= \mathbb{E}N_1 - \mathcal{H}\left(\boldsymbol{X}_1^{N_0} \middle| \boldsymbol{W}_1^{N_1}\oplus\boldsymbol{X}_1^{N_0}, \boldsymbol{X}_{\mathcal{U}}, U, N_1, N_0\right) \\
&= \mathbb{E}N_1 - \alpha_{\tau(k)}\mathcal{H}\left(X_{\tau(k)}\middle|U\right) - \sum_{l\in\mathcal{B}'}\alpha_l && (43) \\
&= \mathbb{E}N_1 - \alpha_{\tau(k)}\frac{2-\delta}{2}\mathcal{H}\left(\frac{1}{2-\delta}\right) - \sum_{l\in\mathcal{B}'}\alpha_l && (44)
\end{aligned}$$

where (43) follows by Lemma 2.

Let $\delta$ be such that the common message of rate $R_{2c}(\delta)$ is decodable at receiver 2. Once the common message is canceled, then the following private rate is achievable

$$\begin{aligned}
R_{2p}(\delta) &= \mathcal{I}\left(\boldsymbol{X}_1^{N_2}; \boldsymbol{X}_{\mathcal{B}'}, V \middle| \boldsymbol{X}_{\mathcal{U}'}, U, N_2\right) \\
&= \sum_{l\in\mathcal{B}'}\mathsf{P}\left(N_2\ge l\right) + \mathsf{P}\left(N_2\ge\tau(k)\right)\frac{2-\delta}{2}\mathcal{H}\left(\frac{1}{2-\delta}\right). && (45)
\end{aligned}$$

It suffices to show that we can make $(R_1(\delta), R_{2p}(\delta)+R_{2c}(\delta))$ coincide with top maximum-sum-rate point by choosing some $\delta$. Note that

$$\begin{aligned}
R_2(\delta) &= R_{2c}(\delta) + R_{2p}(\delta) \\
&= \mathbb{E}[N_0-N_1]^+ + \beta_{\tau(k)}\frac{2-\delta}{2}\mathcal{H}\left(\frac{1}{2-\delta}\right) + \sum_{l\in\mathcal{B}'}\beta_l\,. && (46)
\end{aligned}$$

Multiplying (46) with $\omega_k$ and adding with (44), and noting that $\omega_k\beta_{\tau(k)} = \alpha_{\tau(k)}$, we have

$$R_1(\delta) + \omega_k R_2(\delta) = \mathbb{E}N_1 + \omega_k \mathbb{E}[N_0 - N_1]^+ + \sum_{l\in\mathcal{B}(\omega_k)} (\omega_k\beta_l - \alpha_l) \tag{47}$$

which is exact the equation of boundary of $H(\omega_k)$. Therefore, $(R_1(\delta), R_2(\delta))$ is on the boundary of $H(\omega_k)$ and as $\delta$ varying from 0 to 1, point $(R_1(\delta), R_2(\delta))$ continuously goes from point $L_k$ to point $L_{k-1}$. There must exist some $\delta^*$ such that $\big(R_1(\delta^*), R_2(\delta^*)\big)$ is the top maximum-sum-rate point.

Case 3: The top maximum-sum-rate point is above $L_b$, *i.e.*, the intersection of line $R_2 = \mathbb{E}N_2$ and curve $\mathcal{L}$ is on the boundary of $H(1)$. The basic idea is to split the message of user 2 into a common message and a private message as in case 2; and regard user 1 as two virtual users in order to exploit rate splitting.

Let user 2 transmit its private message using $\boldsymbol{X}_{\mathcal{B}(1)}$ and transmit its common message using $\boldsymbol{X}_{\mathcal{U}(1)}$. Define random vectors $\boldsymbol{U}, \boldsymbol{V} \in \mathbb{F}_2^q$, where the elements of $\boldsymbol{U}$ are i.i.d. Ber$\,(\delta/2)$, and the elements of $\boldsymbol{V}$ are i.i.d Ber$\,(1 - 1/(2-\delta))$, where $\delta \in [0,1]$. We split the user 1 into two virtual users, with codewords $\boldsymbol{U}$ and $\boldsymbol{V}$, respectively. The transmitted codeword consists of $W_i = \max(U_i, V_i)$, $i = 1, \ldots, q$.

Let receiver 1 decode $\boldsymbol{U}$ first, then decode $\boldsymbol{X}_{\mathcal{U}(1)}$ by removing $\boldsymbol{U}$, and finally decode $\boldsymbol{V}$ by removing $\boldsymbol{X}_{\mathcal{U}(1)}$ further. For fixed $\delta$, following rate triple is achievable at receiver 1:

$$\begin{aligned} R_{1,1}(\delta) &= \mathcal{I}\left(\boldsymbol{W}_1^{N_1} \oplus \boldsymbol{X}_1^{N_0}; \boldsymbol{U} \middle| N_1, N_0\right) \\ R_{2c}(\delta) &= \mathcal{I}\left(\boldsymbol{W}_1^{N_1} \oplus \boldsymbol{X}_1^{N_0}; \boldsymbol{X}_{\mathcal{U}(1)} \middle| \boldsymbol{U}, N_1, N_0\right) \\ R_{1,2}(\delta) &= \mathcal{I}\left(\boldsymbol{W}_1^{N_1} \oplus \boldsymbol{X}_1^{N_0}; \boldsymbol{V} \middle| \boldsymbol{X}_{\mathcal{U}(1)}, \boldsymbol{U}, N_1, N_0\right) . \end{aligned}$$

At receiver 2, the following private rate is achievable:

$$R_{2p}(\delta) = \mathcal{I}\left(\boldsymbol{X}_1^{N_2}; \boldsymbol{X}_{\mathcal{B}(1)} \middle| N_2\right)$$

as long as $\delta$ is chosen such that the rate of the common message satisfies $R_{2c}(\delta) + R_{2p}(\delta) \le \mathbb{E}N_2$, so that the common message can be decoded first at both receivers. It then suffices to show that there exists $\delta^* \in [0,1]$ such that $\big(R_1(\delta^*), R_2(\delta^*)\big) = \big(R_{1,1}(\delta^*) + R_{1,2}(\delta^*), R_{2c}(\delta^*) + R_{2p}(\delta^*)\big)$, coincides with the top maximum-sum-rate point.

By the chain rule,

$$\begin{aligned}
R_1(\delta) + R_2(\delta) &= R_{1,1}(\delta) + R_{1,2}(\delta) + R_{2c}(\delta) + R_{2p}(\delta) \\
&= \mathcal{I}\left(\boldsymbol{W}_1^{N_1} \oplus \boldsymbol{X}_1^{N_0}; \boldsymbol{U}, \boldsymbol{V}, \boldsymbol{X}_{\mathcal{U}(1)} \middle| N_1, N_0\right) + \mathcal{I}\left(\boldsymbol{X}_1^{N_2}; \boldsymbol{X}_{\mathcal{B}(1)} \middle| N_2\right) \\
&= \mathcal{I}\left(\boldsymbol{W}_1^{N_1} \oplus \boldsymbol{X}_1^{N_0}; \boldsymbol{W}, \boldsymbol{X}_{\mathcal{U}(1)} \middle| N_1, N_0\right) + \mathcal{I}\left(\boldsymbol{X}_1^{N_2}; \boldsymbol{X}_{\mathcal{B}(1)} \middle| N_2\right) \\
&= \mathcal{I}\left(\boldsymbol{W}_1^{N_1} \oplus \boldsymbol{X}_1^{N_0}; \boldsymbol{W} \middle| N_1, N_0\right) + \mathcal{I}\left(\boldsymbol{W}_1^{N_1} \oplus \boldsymbol{X}_1^{N_0}; \boldsymbol{X}_{\mathcal{U}(1)} \middle| \boldsymbol{W}, N_1, N_0\right) \\
&\qquad\qquad + \mathcal{I}\left(\boldsymbol{X}_1^{N_2}; \boldsymbol{X}_{\mathcal{B}(1)} \middle| N_2\right) \\
&= \mathbb{E}[N_1 - N_0]^+ + \sum_{l \in \mathcal{U}(1)} \mathsf{P}\left(N_0 \geq l\right) + \sum_{l \in \mathcal{B}(1)} \mathsf{P}\left(N_2 \geq l\right) \\
&= \mathbb{E}[N_1 - N_0]^+ + \mathbb{E}N_0 - \sum_{l \in \mathcal{B}(1)} \mathsf{P}\left(N_0 \geq l\right) + \sum_{l \in \mathcal{B}(1)} \mathsf{P}\left(N_2 \geq l\right) \\
&= \mathbb{E}N_1 + \mathbb{E}[N_0 - N_1]^+ + \sum_{l \in \mathcal{B}(1)} \left(\beta_l - \alpha_l\right). \qquad (48)
\end{aligned}$$

Thus, $(R_1(\delta), R_2(\delta))$ is on the boundary of $H(1)$. Note that $R_2(\delta)$ increases as $\delta$ increases because larger $\delta$ indicates larger part of user 1's signal is removed before decoding the common message at receiver 1. Let $\delta = 1$,

$$\begin{aligned}
R_2(1) &= \sum_{l \in \mathcal{U}(1)} \mathsf{P}\left(N_0 \geq l\right) + \sum_{l \in \mathcal{B}(1)} \mathsf{P}\left(N_2 \geq l\right) \\
&\geq \sum_{l=1}^{q} \mathsf{P}\left(N_2 \geq l\right) \\
&= \mathbb{E}N_2
\end{aligned}$$

where the last inequality follows by the fact $\mathsf{P}\left(N_0 \geq l\right) \geq \mathsf{P}\left(N_2 \geq l\right)$ on $l \in \mathcal{U}(1)$. Since both $R_1(\delta)$ and $R_2(\delta)$ are continuous function, there must exist a $\delta^*$ such that $(R_1(\delta^*), R_2(\delta^*))$ is the top maximum-sum-rate point $L_b$, which falls on the line segment between $\big(R_1(0), R_2(0)\big)$ and $\big(R_1(1), R_2(1)\big)$. This completes the proof of Theorem 2.

### *C. Examples*

Before the end of this section, we investigate some special cases for the layered erasure channel with one-sided interference.

*1) The Case of Stochastically Strong Interference:* $\mathsf{P}(N_0 \geq l) \geq \mathsf{P}(N_2 \geq l)$ for every $l \in \{1, \ldots, q\}$. This implies that $\omega\beta_l \leq \alpha_l$ for every $l \in \{1, \ldots, q\}$ and every $\omega \in [0, 1]$. Therefore, the second constraint of (6) can be simplified as $H(0) \cap H(1)$. Hence, the capacity region with stochastically strong interference can be simplified to

$$\left\{ (R_1, R_2) \left| \begin{array}{l} 0 \leq R_1 \leq \mathbb{E}N_1 \\ 0 \leq R_2 \leq \mathbb{E}N_2 \\ R_1 + R_2 \leq \mathbb{E}\max(N_1, N_0) \end{array} \right. \right\}. \tag{49}$$

This generalizes Theorems 3-5 in [18]. We also notice that this result has been essentially established in [24] as a special case.

*2) The Case of Stochastically Weak Interference:* $\mathsf{P}(N_0 \geq l) \leq \mathsf{P}(N_2 \geq l)$ for every $l \in \{1, \ldots, q\}$. Therefore, $\beta_l \geq \alpha_l$ for all $l \in \{1, \ldots, q\}$. The capacity region can be represented by

$$\left\{ (R_1, R_2) \left| \begin{array}{l} 0 \leq R_1 \leq \mathbb{E}N_1 \\ 0 \leq R_2 \leq \mathbb{E}N_2 \\ R_1 + \omega_k R_2 \leq \mathbb{E}N_1 + \omega_k \mathbb{E}(N_0 - N_1)^+ \\ \qquad\qquad + \sum_{l=1}^{q} (\omega_k \beta_l - \alpha_l)^+ \quad k \in \{1, \ldots, q\} \end{array} \right. \right\}. \tag{50}$$

Furthermore, the sum capacity is

$$\begin{aligned} C_{sum} &= \mathbb{E}N_1 + \mathbb{E}[N_0 - N_1]^+ + \sum_{l=1}^{q} (\beta_l - \alpha_l) \\ &= \mathbb{E}\max(N_0, N_1) + \mathbb{E}N_2 - \mathbb{E}N_0 \end{aligned}$$

which is a generalization of Theorem 7 in [18].

*3) The Case of Pure Deterministic Model:* $N_0 \equiv n_0$, $N_1 \equiv n_1$, and $N_2 \equiv n_2$ are known to both the transmitter and the receiver. If $n_0 \geq n_2$, it falls into the case of stochastically strong interference. If $n_0 \leq n_2$, it falls into the case of stochastically weak interference where $\omega(l) = 1$ for $l \in \{1, \ldots, (n_2 - n_0)\}$ and $\omega(l) = 0$ otherwise. Therefore, by simplifying (49) and (50), the capacity region becomes

$$\left\{ (R_1, R_2) \left| \begin{array}{l} 0 \leq R_1 \leq n_1 \\ 0 \leq R_2 \leq n_2 \\ R_1 + R_2 \leq \max(n_1, n_0) + (n_2 - n_1)^+ \end{array} \right. \right\} \tag{51}$$

which follows previous results on the capacity region of deterministic interference channels in [6], [25].

## VI. One-sided Fading Gaussian Interference Channel

In this section, we study the one-sided fading Gaussian interference channel model described in Section II and prove Theorem 1. An outer bound for the capacity region is first derived by converting the channel to a "layered" model through the use of "incremental channels." An achievability result is then developed in analogy to the coding techniques introduced for the layered erasure model. The capacity for a few special cases is provided at the end of this section.

### *A. The Outer Bound for the Gaussian Model*

Note that the capacity region depends only on the marginal distributions of the channel outputs $Y$ and $Z$ conditioned on the channel inputs and states. We can thus replace the joint distribution by any distribution compatible with identical conditional marginals, so that the capacity region is preserved. Throughout Section VI-A, we assume alignment of the links as follows without changing the capacity region. First, because the phases are known and can be compensated at the receivers, we can assume $\Theta_{0m} = \Theta_{2m} = 0$ for all $m = 1, \ldots, n$ without loss of generality.

Secondly, we assume that the fading states of the direct link for user 2 and the interference link are aligned, *i.e.*, for every $m = 1, \ldots, n$, their SNRs are driven by the same random variable $\Lambda_m$:

$$S_{0m} = F_{S_0}^{-1}(\Lambda_m), \quad \text{and} \quad S_{2m} = F_{S_2}^{-1}(\Lambda_m)\,,$$

where $\{\Lambda_m\}$ are i.i.d. and uniformly distributed on $[0, 1]$. The states $\{S_{1m}\}$ remain independent of $\{S_{0m}, S_{2m}\}$. It is important to note that the region $\overline{\mathcal{R}}$ remains the same, because the bounds in (3) are invariant to the dependence of $S_0$ and $S_2$ introduced here.

Third, we assume that the additive noises at the two receivers are also aligned. This is easy because Gaussian noise is infinitely divisible. Let $\{B_m(\nu), \nu \geq 0\}$, $m = 1, \ldots, n$ be $n$ independent CSCG continuous-time processes, each of which is of independent increments with $\mathbb{E}\{|B_m(\nu)|^2\} = \nu$. Basically $B_m$ is a complex-valued Brownian motion. Without changing the capacity region, we can simplify the model (1) to the following:

$$\begin{aligned} Y_m &= X_m + \sqrt{\frac{S_1}{S_0}}\, e^{j\Theta_m} W_m + B_m\left(\frac{1}{S_0}\right) \\ Z_m &= X_m + B_m\left(\frac{1}{S_2}\right) \end{aligned}$$

where $m = 1, \ldots, n$ and $\Theta_m$ are i.i.d. uniform on $[0, 2\pi]$.

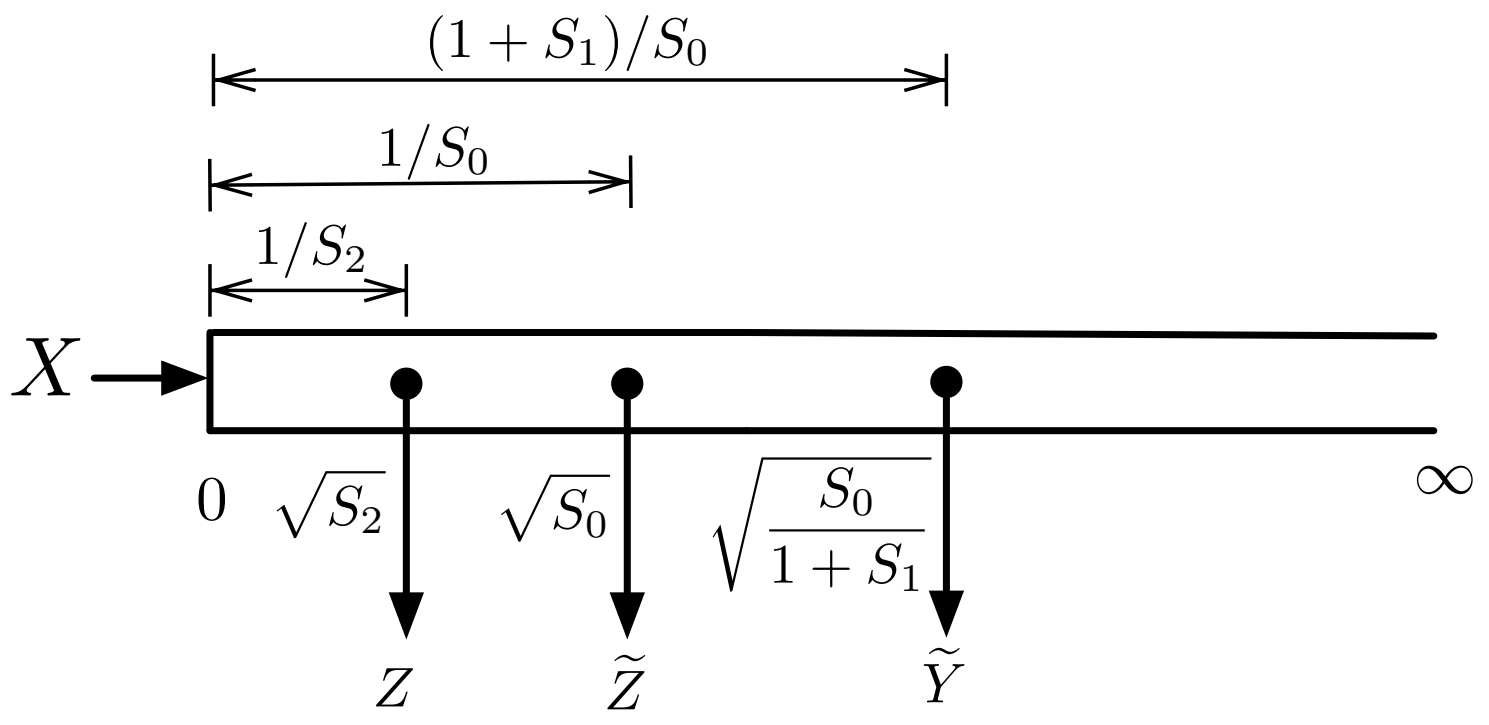

Fig. 4. An illustration of noise alignment via incremental channel. The signal $X$ is corrupted by a circularly symmetric standard complex Brownian motion. $Z$, $\widetilde{Z}$, and $\widetilde{Y}$ are generated by taking the corrupted signal out at time $1/S_2$, $1/S_0$, and $(S_1+1)/S_0$, respectively.

For convenience, let $\boldsymbol{S} = \big(S_0, S_1, S_2\big)$ and $\boldsymbol{\Theta} = \big(\Theta_0, \Theta_1, \Theta_2\big)$. The notation for the time index in this section is different than that used for the layered erasure model in Sections IV and V: In general $X_m$ refers to a signal at time interval $m$, and $\boldsymbol{X}^n$ refers to the signal over $n$ time intervals, $(X_1, \ldots, X_n)$. Moreover, let $\boldsymbol{S}^n = \{S_i[j] \,|\, i = 0, 1, 2 \text{ and } j = 1, 2, \ldots, n\}$ and $\boldsymbol{\Theta}^n = \{\Theta_i[j] \,|\, i = 0, 1, 2 \text{ and } j = 1, 2, \ldots, n\}$.

By Fano's inequality, the rates of the two users must satisfy

$$nR_1 - n\delta_n \le \mathcal{I}\left(\boldsymbol{W}^n; \boldsymbol{Y}^n | \boldsymbol{S}^n, \boldsymbol{\Theta}^n\right) \tag{52}$$

$$nR_2 - n\delta_n \le \mathcal{I}\left(\boldsymbol{X}^n; \boldsymbol{Z}^n | \boldsymbol{S}^n\right) \ . \tag{53}$$

For convenience, let us introduce signals $\widetilde{Z}$ and $\widetilde{Y}$ as follows:

$$\begin{aligned}
\widetilde{Z} &= X + B\left(\frac{1}{S_0}\right) \\
\widetilde{Y} &= X + \sqrt{\frac{S_1}{S_0}}\,\widetilde{W} + B\left(\frac{1}{S_0}\right) \\
&= X + B\left(\frac{S_1+1}{S_0}\right)
\end{aligned}$$

where we have implicitly defined $\widetilde{W}$ as a unit CSCG random variable, which is proportional to the increment of the Brownian motion, and is hence independent of $X$ and the additive noise $B(1/S_0)$. By [26, Corollary 2], setting the distribution of the input $W$ to unit CSCG incurs no more than 1 bit of loss in the mutual information on the RHS of (52). That is

$$nR_1 - n\delta_n \le \mathcal{I}\left(\widetilde{\boldsymbol{W}}^n; \widetilde{\boldsymbol{Y}}^n \middle| \boldsymbol{S}^n\right) + n \ .$$

Because $\widetilde{W}$—$(\widetilde{Y} - X)$—$\widetilde{Y}$ is a Markov chain,

$$\begin{aligned} nR_1 - n\delta_n - n &\le \mathcal{I}\left(\widetilde{\boldsymbol{W}}^n; \widetilde{\boldsymbol{Y}}^n | \boldsymbol{S}^n\right) \\ &= \mathcal{I}\left(\widetilde{\boldsymbol{W}}^n; \widetilde{\boldsymbol{Y}}^n - \boldsymbol{X}^n \middle| \boldsymbol{S}^n\right) - \mathcal{I}\left(\widetilde{\boldsymbol{W}}^n; \widetilde{\boldsymbol{Y}}^n - \boldsymbol{X}^n \middle| \widetilde{\boldsymbol{Y}}^n, \boldsymbol{S}^n\right) \\ &= n\mathbb{E}\log\left(1 + S_1\right) - \mathcal{I}\left(\widetilde{\boldsymbol{Z}}^n; \boldsymbol{X}^n \middle| \widetilde{\boldsymbol{Y}}^n, \boldsymbol{S}^n\right) \ . \end{aligned} \tag{54}$$

Meanwhile, the bound (53) on the rate of user 2 becomes

$$\begin{aligned} nR_2 - n\delta_n &\le \mathcal{I}\left(\boldsymbol{X}^n; \boldsymbol{Z}^n | \boldsymbol{S}^n\right) \\ &\le \mathcal{I}\left(\boldsymbol{X}^n; \boldsymbol{Z}^n, \widetilde{\boldsymbol{Y}}^n \middle| \boldsymbol{S}^n\right) \\ &= \mathcal{I}\left(\boldsymbol{X}^n; \widetilde{\boldsymbol{Y}}^n \middle| \boldsymbol{S}^n\right) + \mathcal{I}\left(\boldsymbol{X}^n; \boldsymbol{Z}^n \middle| \widetilde{\boldsymbol{Y}}^n, \boldsymbol{S}^n\right) \\ &\le n\mathbb{E}\log\left(1 + \frac{S_0}{1 + S_1}\right) + \mathcal{I}\left(\boldsymbol{X}^n; \boldsymbol{Z}^n \middle| \widetilde{\boldsymbol{Y}}^n, \boldsymbol{S}^n\right) \ . \end{aligned} \tag{55}$$

We have the following result, which is a parallel of Lemma 2.

*Lemma 3:* Suppose a collection of random variables $\boldsymbol{T}$ is independent of $\boldsymbol{S} = (S_0, S_1, S_2)$. Then

$$\mathcal{I}\left(X; \widetilde{Z} \middle| \widetilde{Y}, \boldsymbol{T}, \boldsymbol{S}\right) = \log e \int_0^\infty \alpha(\gamma) \mathsf{mmse}\left[X | \gamma, \boldsymbol{T}\right] \mathsf{d}\gamma \tag{56}$$

$$\mathcal{I}\left(X; Z \middle| \widetilde{Y}, \boldsymbol{T}, \boldsymbol{S}\right) = \log e \int_0^\infty \beta(\gamma) \mathsf{mmse}\left[X | \gamma, \boldsymbol{T}\right] \mathsf{d}\gamma \tag{57}$$

where $\alpha(\gamma)$ and $\beta(\gamma)$ are given in (4) and (5), respectively, and $\mathsf{mmse}\left[X|\gamma, \boldsymbol{T}\right]$ is defined as

$$\mathsf{mmse}\left[X|\gamma, \boldsymbol{T}\right] = \mathbb{E}\left[X - \mathbb{E}[X|\sqrt{\gamma}X + U, \boldsymbol{T}]\right]^2 .$$

The proof of Lemma 3 is given in Appendix II and a similar interpretation of the result as the one for Lemma 2 can be obtained. In order to establish the third constraint in (3) in Theorem 1, we consider the weighted difference between the two mutual informations in (54) and (55):

$$\begin{aligned}
&\omega\mathcal{I}\left(\boldsymbol{X}^n;\boldsymbol{Z}^n\middle|\widetilde{\boldsymbol{Y}}^n,\boldsymbol{S}^n\right)-\mathcal{I}\left(\boldsymbol{X}^n;\widetilde{\boldsymbol{Z}}^n\middle|\widetilde{\boldsymbol{Y}}^n,\boldsymbol{S}^n\right)\\
&\quad=\mathcal{I}\left(\boldsymbol{X}^n;\boldsymbol{Z}^n\middle|\widetilde{\boldsymbol{Y}}^n,\boldsymbol{S}^n\right)-\mathcal{I}\left(\boldsymbol{X}^n;\widetilde{\boldsymbol{Z}}^n\middle|\widetilde{\boldsymbol{Y}}^n,\boldsymbol{S}^n\right)+(\omega-1)\mathcal{I}\left(\boldsymbol{X}^n;\boldsymbol{Z}^n\middle|\widetilde{\boldsymbol{Y}}^n,\boldsymbol{S}^n\right)\\
&\quad=\sum_{m=1}^{n}\left[\mathcal{I}\left(\boldsymbol{X}^n;\boldsymbol{Z}^m,\widetilde{\boldsymbol{Z}}^n_{m+1}\middle|\widetilde{\boldsymbol{Y}}^n,\boldsymbol{S}^n\right)-\mathcal{I}\left(\boldsymbol{X}^n;\boldsymbol{Z}^{m+1},\widetilde{\boldsymbol{Z}}^n_{m}\middle|\widetilde{\boldsymbol{Y}}^n,\boldsymbol{S}^n\right)\right]\\
&\qquad\qquad+(\omega-1)\mathcal{I}\left(\boldsymbol{X}^n;\boldsymbol{Z}^n\middle|\widetilde{\boldsymbol{Y}}^n,\boldsymbol{S}^n\right) \qquad (58)
\end{aligned}$$

by "Marton-like" expansion as for the layered erasure model. By the chain rule, the RHS of (58) can be reduced to:

$$\begin{aligned}
&\sum_{m=1}^{n}\left[\mathcal{I}\left(Z_m;\boldsymbol{X}^n\middle|\boldsymbol{Z}^{m-1},\widetilde{\boldsymbol{Z}}^n_{m+1},\widetilde{\boldsymbol{Y}}^n,\boldsymbol{S}^n\right)-\mathcal{I}\left(\widetilde{Z}_m;\boldsymbol{X}^n\middle|\boldsymbol{Z}^{m-1},\widetilde{\boldsymbol{Z}}^n_{m+1},\widetilde{\boldsymbol{Y}}^n,\boldsymbol{S}^n\right)\right]\\
&\qquad\qquad+(\omega-1)\sum_{m=1}^{n}\mathcal{I}\left(Z_m;\boldsymbol{X}^n\middle|\boldsymbol{Z}^{m-1},\widetilde{\boldsymbol{Y}}^n,\boldsymbol{S}^n\right)\\
&\quad\le\sum_{m=1}^{n}\left[\mathcal{I}\left(Z_m;\boldsymbol{X}^n\middle|\boldsymbol{Z}^{m-1},\widetilde{\boldsymbol{Z}}^n_{m+1},\widetilde{\boldsymbol{Y}}^n,\boldsymbol{S}^n\right)-\mathcal{I}\left(\widetilde{Z}_m;\boldsymbol{X}^n\middle|\boldsymbol{Z}^{m-1},\widetilde{\boldsymbol{Z}}^n_{m+1},\widetilde{\boldsymbol{Y}}^n,\boldsymbol{S}^n\right)\right]\\
&\qquad\qquad+(\omega-1)\sum_{m=1}^{n}\mathcal{I}\left(Z_m;\boldsymbol{X}^n\middle|\boldsymbol{Z}^{m-1},\widetilde{\boldsymbol{Z}}^n_{m+1},\widetilde{\boldsymbol{Y}}^n,\boldsymbol{S}^n\right) \qquad (59)\\
&\quad=\sum_{m=1}^{n}\left[\omega\mathcal{I}\left(Z_m;X_m\middle|\boldsymbol{Z}^{m-1},\widetilde{\boldsymbol{Z}}^n_{m+1},\widetilde{\boldsymbol{Y}}^n,\boldsymbol{S}^n\right)-\mathcal{I}\left(\widetilde{Z}_m;X_m\middle|\boldsymbol{Z}^{m-1},\widetilde{\boldsymbol{Z}}^n_{m+1},\widetilde{\boldsymbol{Y}}^n,\boldsymbol{S}^n\right)\right]. \qquad (60)
\end{aligned}$$

where (59) is due to the fact that $Z_m$—$\boldsymbol{X}^n$—$\widetilde{\boldsymbol{Z}}^n_{m+1}$ is Markovian and $\omega \le 1$, and (60) is because that $Z_m$—$X_m$—$(\boldsymbol{X}^{m-1},\boldsymbol{X}^n_{m+1})$ and $\widetilde{Z}_m$—$X_m$—$(\boldsymbol{X}^{m-1},\boldsymbol{X}^n_{m+1})$ are both Markovian. Finally, for each $m$, we apply Lemma 3 with

$$\boldsymbol{T}_m=\left(\boldsymbol{Z}^{m-1},\widetilde{\boldsymbol{Z}}^n_{m+1},\widetilde{\boldsymbol{Y}}^{m-1},\widetilde{\boldsymbol{Y}}^n_{m+1},\boldsymbol{S}^{m-1},\boldsymbol{S}^n_{m+1}\right)$$

which is independent of $(S_{0m},S_{1m},S_{2m})$, to obtain

$$\omega\mathcal{I}\left(\boldsymbol{Z}^n;\boldsymbol{X}^n\middle|\widetilde{\boldsymbol{Y}}^n,\boldsymbol{S}^n\right)-\mathcal{I}\left(\widetilde{\boldsymbol{Z}}^n;\boldsymbol{X}^n\middle|\widetilde{\boldsymbol{Y}}^n,\boldsymbol{S}^n\right)\le\sum_{m=1}^{n}\log e\int_0^\infty(\omega\beta(\gamma)-\alpha(\gamma))\,\mathsf{mmse}\left[X_m\middle|\boldsymbol{T}_m\right]\mathsf{d}\gamma\,.$$

Since

$$\mathsf{mmse}\left[X_m|\gamma,\boldsymbol{T}_m\right]\le\frac{1}{1+\gamma}\quad\forall\gamma,$$

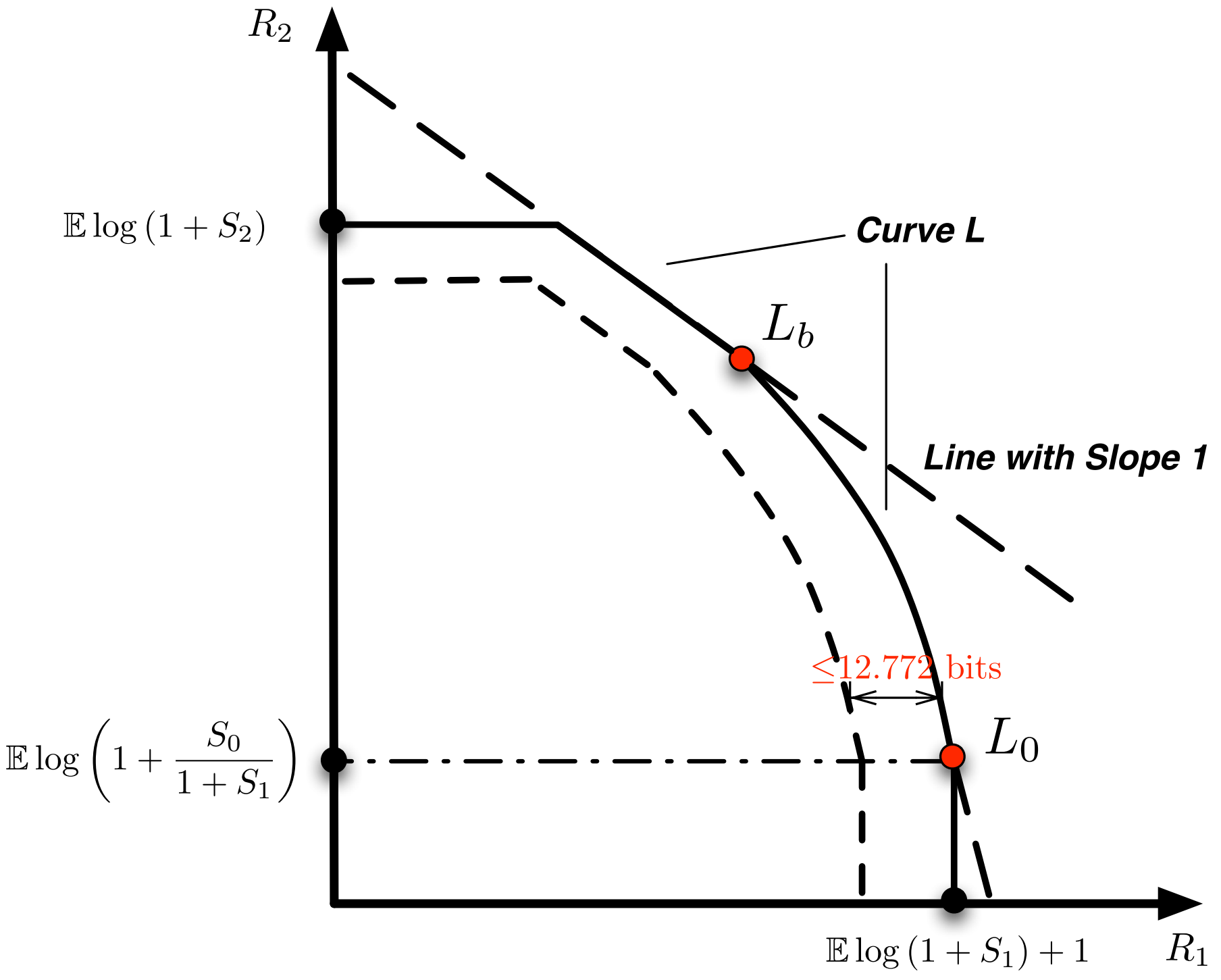

Fig. 5. An illustration of the inner and outer bounds for capacity region. The outer bound is drawn in solid line; the inner bound is drawn in dashed line. Note that we drop the first constraint in the (3) so that the $R_1 \leq \mathbb{E}\log(1+S_1)+1$.

we have

$$\omega\mathcal{I}\left(\boldsymbol{Z}^n;\boldsymbol{X}^n\middle|\widetilde{\boldsymbol{Y}}^n,\boldsymbol{S}^n\right)-\mathcal{I}\left(\widetilde{\boldsymbol{Z}}^n;\boldsymbol{X}^n\middle|\widetilde{\boldsymbol{Y}}^n,\boldsymbol{S}^n\right)\leq n\int_0^\infty(\omega\beta(\gamma)-\alpha(\gamma))^+\frac{\log e}{1+\gamma}\mathsf{d}\gamma\,. \tag{61}$$

Comparing (54), (55) and (61) and noting that $\delta_n \to 0$ as $n \to \infty$, we have established the third constraint in (3).

### B. The Inner Bound: A Constant Gap Result

We propose a coding scheme which achieves a rate region within a constant gap to the outer bound developed in Section VI-A. The gap applies to all SNR and fading statistics. Thus the inner and outer bounds are asymptotically tight at high SNRs, where the capacity becomes large. The coding scheme is inspired by the coding scheme used for fading broadcast channels developed by Tse and Yates [17].

To make the analysis easier, we drop the first constraint in (3) and the new region is still an outer bound, denoted by $\overline{\mathcal{R}}'$. Note that the third bound at $\omega = 0$ corresponds to $R_1 \leq \mathbb{E}\log(1+S_1)+1$, which is looser than the first constraint in (3), but within 1 bit.

Similar to the capacity region for the layered erasure model, besides the two axes, the outer bound $\overline{\mathcal{R}}'$ is enclosed by two curves, which correspond the remaining two constraints: One is line $R_2 = \mathbb{E}\log(1+S_2)$; the other curve $\mathcal{L}$ is the boundary of the region $\cap_{\omega\in[0,1]} H(\omega)$, where

$$H(\omega) = \left\{(R_1,R_2)\middle| R_1+\omega R_2 \leq 1+\mathbb{E}\log(1+S_1)+\omega\mathbb{E}\log\left(1+\frac{S_0}{1+S_1}\right) \right. \\ \left. + \int_{\mathcal{B}(\omega)} \big(\omega\beta(\gamma)-\alpha(\gamma)\big)\mathsf{d}\gamma \right\}$$

and

$$\mathcal{B}(\omega) = \{\gamma\in[0,\infty) | \omega\beta(\gamma) \geq \alpha(\gamma)\}.$$

We claim that for every $\omega\in[0,1]$, the straight line boundary of $H(\omega)$ touches the curve $\mathcal{L}$ at the point

$$\begin{aligned}&\big(\overline{R}_1(\omega),\overline{R}_2(\omega)\big) \\ &= \left(1+\mathbb{E}\log(1+S_1)-\int_{\mathcal{B}(\omega)}\alpha(\gamma)\mathsf{d}\gamma\ ,\quad \mathbb{E}\log\left(1+\frac{S_0}{1+S_1}\right)+\int_{\mathcal{B}(\omega)}\beta(\gamma)\mathsf{d}\gamma\right). \qquad (62)\end{aligned}$$

To see this, first, note that the point is on the boundary of $H(\omega)$ because it achieves the equality of the constraint $H(\omega)$. Moreover, for every $\omega'\in[0,1]$, we have

$$\begin{aligned}\overline{R}_1(\omega)+\omega'\overline{R}_2(\omega) &= 1+\mathbb{E}\log(1+S_1)+\omega'\mathbb{E}\log\left(1+\frac{S_0}{1+S_1}\right)+\int_{\mathcal{B}(\omega)}\big(\omega'\beta(\gamma)-\alpha(\gamma)\big)\mathsf{d}\gamma \\ &\leq 1+\mathbb{E}\log(1+S_1)+\omega'\mathbb{E}\log\left(1+\frac{S_0}{1+S_1}\right)+\int_{\mathcal{B}(\omega')}\big(\omega'\beta(\gamma)-\alpha(\gamma)\big)\mathsf{d}\gamma\end{aligned}$$

where the last step is due to the fact that $\omega'\beta(\gamma)-\alpha(\gamma)\leq 0$ for every $\gamma\in\mathcal{B}(\omega)\backslash\mathcal{B}(\omega')$. Thus $\big(\overline{R}_1(\omega),\overline{R}_2(\omega)\big)\in H(\omega')$, which proves the claim.

Denote points $\big(\overline{R}_1(1),\overline{R}_2(1)\big)$ and $\big(\overline{R}_1(0),\overline{R}_2(0)\big)$ by $L_b$ and $L_0$, respectively. Generally, the curve $\mathcal{L}$ can be divided into three parts: The part on the left side of $L_b$ is a ray with slope -1; the part between $L_b$ and $L_0$ has tangent line with slope steeper than -1; The remaining part is a vertically downward ray starting from point $L_0$. Another observation is that all the extreme points are contained in the closure set $\mathcal{M} = \overline{\left\{\big(\overline{R}_1(\omega),\overline{R}_2(\omega)\big)\middle|\omega\in[0,1]\right\}}$. However, not all points between $L_b$ and $L_0$ are contained in the set $\mathcal{M}$. For example, when $S_0$, $S_1$, $S_2$ are all discrete random variables, the outer bound $\overline{\mathcal{R}}$ becomes a polyhedron, like the case of layered erasure model.

The line $R_2 = \mathbb{E}\log(1+S_2)$ can intersect with curve $\mathcal{L}$ in various locations and we refer to the intersection as the *top maximum sum-rate point* as in the case of the layered erasure model. In the following, we first show that every point in $\mathcal{M}$ and below line $R_2 = \mathbb{E}\log(1+S_2)$ is achievable within a constant gap, and then deal with the top maximum sum-rate point.

The points in $\mathcal{M}$ can either be parametrized with $\omega$ as (62) or be asymptotically approached by those can be parametrized with $\omega$. Therefore, it suffices to show the achievability result for those can be parametrized. Note that $\mathsf{P}(S_2 \geq \gamma) \geq \mathsf{P}(S_0 \geq \gamma)$ whenever $\gamma \in \mathcal{B}(\omega)$. Thus, the coordinate (62) can be rewritten as

$$\overline{R}_1(\omega) = 1 + \mathbb{E}\log(1+S_1) + \int_{\mathcal{B}(\omega)} \mathsf{P}\left(\frac{S_0}{S_1+1} \geq \gamma\right)\frac{\log e}{1+\gamma}\mathsf{d}\gamma - \int_{\mathcal{B}(\omega)} \mathsf{P}(S_0 \geq \gamma)\frac{\log e}{1+\gamma}\mathsf{d}\gamma \tag{63a}$$

$$\overline{R}_2(\omega) = \int_{\mathcal{B}^c(\omega)} \mathsf{P}\left(\frac{S_0}{S_1+1} \geq \gamma\right)\frac{\log e}{1+\gamma}\mathsf{d}\gamma + \int_{\mathcal{B}(\omega)} \mathsf{P}(S_2 \geq \gamma)\frac{\log e}{1+\gamma}\mathsf{d}\gamma\,. \tag{63b}$$

In following, we show that the rate pair $\left(\overline{R}_1(\omega) - \Delta^*, \overline{R}_2(\omega) - \Delta^*\right)$ can be achieved for some universal constant $\Delta^*$.

Let the transmitter 1 generate a random codebook with i.i.d. unit CSCG distribution. Let the signaling of user 2 follow the distribution of $X$, which is constructed as follows

$$X = \sqrt{\frac{3}{2}}\sum_{i=1}^{\infty}\widetilde{X}_{Ii}2^{-i} + j\sqrt{\frac{3}{2}}\sum_{i=1}^{\infty}\widetilde{X}_{Qi}2^{-i} \tag{64}$$

where $\{\widetilde{X}_{Ii}\}_1^{\infty}$ and $\{\widetilde{X}_{Qi}\}_1^{\infty}$ are independent signals taking $\pm 1$ equally likely. Let $X_{\mathcal{N}}$ denote $\sqrt{3/2}\sum_{i\in\mathcal{N}}\left(\widetilde{X}_{Ii} + j\widetilde{X}_{Qi}\right)$. Fix $\rho > 0$ to be a constant, and let $\rho_n = \rho 2^{2(n-1)}$, $n = 1, 2, \ldots$. For $\omega \in [0,1]$, let $\mathcal{N}(\omega) = \{n \geq 1 : \rho_n \in \mathcal{B}(\omega)\}$ and $\mathcal{N}^c(\omega) = \{1, 2, \ldots\}\backslash\mathcal{N}(\omega)$. The transmitted signal of user 2 is the sum of two codewords, $X[m] = X_{\mathcal{N}(\omega)}[m] + X_{\mathcal{N}^c(\omega)}[m]$, $m = 1, \ldots, n$, which carry the common message and the private message respectively. The codebooks for the common message and the private message are randomly generated using the distributions of $X_{\mathcal{N}^c(\omega)}$ and $X_{\mathcal{N}(\omega)}$, respectively.

Receiver 1 first decodes the common message of user 2 and then decodes its own message.

Therefore user 2 can achieve rate

$$
\begin{aligned}
R_2(\omega) &= \min\Big(\mathcal{I}\left(\sqrt{S_1}\,e^{j\Theta_1}W + \sqrt{S_0}\,e^{j\Theta_0}X + U; X_{\mathcal{N}^c(\omega)}\Big|S_1, S_0, \Theta_1, \Theta_0\right) \\
&\qquad + \mathcal{I}\left(\sqrt{S_2}\,e^{j\Theta_2}X + V; X_{\mathcal{N}(\omega)}\Big|S_2, \Theta_2\right)\,,\ \mathcal{I}\left(\sqrt{S_2}\,e^{j\Theta_2}X + V; X\Big|S_2, \Theta_2\right)\Big) \\
&= \min\Big(\mathcal{I}\left(\sqrt{S_1}\,W + \sqrt{S_0}\,X + U; X_{\mathcal{N}^c(\omega)}\Big|S_1, S_0\right) \\
&\qquad + \mathcal{I}\left(\sqrt{S_2}\,X + V; X_{\mathcal{N}(\omega)}\Big|S_2\right)\,,\ \mathcal{I}\left(\sqrt{S_2}\,X + V; X\Big|S_2\right)\Big) \qquad (65)
\end{aligned}
$$

where the phase random variables are removed in (65). The reason is following: $\Theta_1$ is absorbed by $W$ since it is CSCG; $\Theta_0$ and $\Theta_2$ can be compensated at receivers since receivers know each realization of them. Similarly, we can remove the phase random variables in the remaining development. After removing the common message, receiver 1 can decode its own message at rate

$$
\begin{aligned}
R_1(\omega) &= \mathcal{I}\left(\sqrt{S_1}\,e^{j\Theta_1}W + \sqrt{S_0}\,e^{j\Theta_0}X + U; W\Big|X_{\mathcal{N}^c(\omega)}, S_1, S_0, \Theta_1, \Theta_0\right) \\
&= \mathcal{I}\left(\sqrt{S_1}\,W + \sqrt{S_0}\,X_{\mathcal{N}(\omega)} + U; W\Big|S_1, S_0\right) \\
&= \mathcal{I}\left(\sqrt{S_1}\,W + U; W\Big|S_1\right) \\
&\qquad - \mathcal{I}\left(\sqrt{S_1}\,W + U; W\Big|\sqrt{S_1}\,W + \sqrt{S_0}\,X_{\mathcal{N}(\omega)} + U, S_0, S_1\right) \qquad (66) \\
&= \mathbb{E}\log\left(1 + S_1\right) - \mathcal{I}\left(\sqrt{S_0}\,X_{\mathcal{N}(\omega)} + U; X_{\mathcal{N}(\omega)}\Big|\sqrt{S_1}\,W + \sqrt{S_0}\,X_{\mathcal{N}(\omega)} + U, S_0, S_1\right) \\
&= \mathbb{E}\log\left(1 + S_1\right) - \mathcal{I}\left(\sqrt{S_0}\,X_{\mathcal{N}(\omega)} + U; X_{\mathcal{N}(\omega)}\Big|S_0\right) \\
&\qquad\qquad + \mathcal{I}\left(\sqrt{S_1}\,W + \sqrt{S_0}\,X_{\mathcal{N}(\omega)} + U; X_{\mathcal{N}(\omega)}\Big|S_1, S_0\right)\,. \qquad (67)
\end{aligned}
$$

where (66) and (67) are due to the fact that $W$—$(\sqrt{S_1}\,W + U)$—$(\sqrt{S_1}\,W + \sqrt{S_0}\,X_{\mathcal{N}} + U)$ and $X_{\mathcal{N}}$—$(\sqrt{S_0}\,X_{\mathcal{N}} + U)$—$(\sqrt{S_1}\,W + \sqrt{S_0}\,X_{\mathcal{N}} + U)$ are Markovian.

Now, we compare $\left(\overline{R}_1(\omega), \overline{R}_2(\omega)\right)$ with $\left(R_1(\omega), R_2(\omega)\right)$. By (63a) and (67),

$$
\begin{aligned}
&\overline{R}_1(\omega) - R_1(\omega) \\
&\quad = 1 + \left[\int_{\mathcal{B}(\omega)} \mathsf{P}\left(\frac{S_0}{S_1+1} \geq \gamma\right)\frac{\log e}{1+\gamma}\mathsf{d}\gamma - \mathcal{I}\left(\sqrt{S_1}\,W + \sqrt{S_0}\,X_{\mathcal{N}(\omega)} + U; X_{\mathcal{N}(\omega)}\Big|S_1, S_0\right)\right] \\
&\qquad + \left[\mathcal{I}\left(\sqrt{S_0}\,X_{\mathcal{N}(\omega)} + U; X_{\mathcal{N}(\omega)}\Big|S_0\right) - \int_{\mathcal{B}(\omega)} \mathsf{P}\left(S_0 \geq \gamma\right)\frac{\log e}{1+\gamma}\mathsf{d}\gamma\right] \\
&\quad \leq 1 + \left[\int_{\mathcal{B}(\omega)} \mathsf{P}\left(\frac{S_0}{S_1+1} \geq \gamma\right)\frac{\log e}{1+\gamma}\mathsf{d}\gamma - \mathcal{I}\left(\sqrt{S_1}\,W + \sqrt{S_0}\,X + U; X_{\mathcal{N}(\omega)}\Big|S_1, S_0\right)\right]
\end{aligned}
$$

$$+\left[\mathcal{I}\left(\sqrt{S_0}\,X_{\mathcal{N}(\omega)}+U;X_{\mathcal{N}(\omega)}\middle|S_0\right)-\int_{\mathcal{B}(\omega)}\mathsf{P}\left(S_0\geq\gamma\right)\frac{\log e}{1+\gamma}\mathsf{d}\gamma\right] \tag{68}$$

where we replace $X_{\mathcal{N}(\omega)}$ with $X$ in the first mutual information due to the fact that $X_{\mathcal{N}(\omega)}$ and $X_{\mathcal{N}^c(\omega)}$ are mutually independent.

We have following crucial lemma:

*Lemma 4:* Let $Z=\sqrt{\Gamma}\,X+U$, where $\Gamma$ is an arbitrary non-negative random variable. Let $\mathcal{B}$ be a measurable subset of $[0,\infty)$. We define $\mathcal{N}=\{n\geq 1|\rho_n\in\mathcal{B}\}$. The following inequalities hold

$$\int_{\mathcal{B}}\mathsf{P}\left(\Gamma\geq\gamma\right)\frac{\log e}{1+\gamma}\mathsf{d}\gamma-\Delta(\rho)\leq\mathcal{I}\left(Z;X_{\mathcal{N}}\middle|\Gamma\right) \tag{69}$$

and

$$\mathcal{I}\left(\sqrt{\Gamma}\,X_{\mathcal{N}}+U;X_{\mathcal{N}}\middle|\Gamma\right)\leq\int_{\mathcal{B}}\mathsf{P}\left(\Gamma\geq\gamma\right)\frac{\log e}{1+\gamma}\mathsf{d}\gamma+\Delta(\rho)-C \tag{70}$$

where $\Delta(\rho)$ is independent of $\Gamma$, and $C=\log\left(\pi e/6\right)+1\approx 1.546$ bit. Furthermore, $\Delta(\rho)$ can be minimized by choosing $\rho=5.65$, yielding $\Delta\leq 6.386$ bit.

Lemma 4 is proved in Appendix III. We note that inequality (69) has basically been shown in [17].

Applying Lemma 4 to (68), we have

$$\begin{aligned}\overline{R}_1(\omega)-R_1(\omega)&\leq 1+\Delta(\rho)+\Delta(\rho)-C\\&\leq 2\Delta(\rho)\,.\end{aligned}$$

By (65),

$$\begin{aligned}&\overline{R}_2(\omega)-R_2(\omega)\\&=\max\Bigg(\overline{R}_2(\omega)-\mathcal{I}\left(\sqrt{S_1}\,W+\sqrt{S_0}\,X+U;X_{\mathcal{N}^c(\omega)}\middle|S_1,S_0\right)-\mathcal{I}\left(\sqrt{S_2}\,X+V;X_{\mathcal{N}(\omega)}\middle|S_2\right)\,,\\&\qquad\qquad\overline{R}_2(\omega)-\mathcal{I}\left(\sqrt{S_2}\,X+V;X\middle|S_2\right)\Bigg)\,.\end{aligned} \tag{71}$$

By (63b), the first term in (71) can be written as

$$\begin{aligned}&\int_{\mathcal{B}^c(\omega)}\mathsf{P}\left(\frac{S_0}{S_1+1}\geq\gamma\right)\frac{\log e}{1+\gamma}\mathsf{d}\gamma-\mathcal{I}\left(\sqrt{S_1}\,W+\sqrt{S_0}\,X+U;X_{\mathcal{N}^c(\omega)}\middle|S_1,S_0\right)\\&\quad+\int_{\mathcal{B}(\omega)}\mathsf{P}\left(S_2\geq\gamma\right)\frac{\log e}{1+\gamma}\mathsf{d}\gamma-\mathcal{I}\left(\sqrt{S_2}\,X+V;X_{\mathcal{N}(\omega)}\middle|S_2\right)\\&\leq\Delta(\rho)+\Delta(\rho)\end{aligned} \tag{72}$$

by Lemma 4. Meanwhile, since $X$ is uniform, by [26, Eq. (10)],

$$\mathbb{E}\log(1+S_2) - \mathcal{I}\left(\sqrt{S_2}\,X+V;X\middle|S_2\right) \leq \log\left(\frac{\pi e}{6}\right)+1\,.$$

By assumption, $\overline{R}_2(\omega)$ is below $\mathbb{E}\log(1+S_2)$, so that

$$\overline{R}_2(\omega) - \mathcal{I}\left(\sqrt{S_2}\,X+V;X\middle|S_2\right) \leq \log\left(\frac{\pi e}{6}\right)+1\,. \tag{73}$$

Putting (71), (72) and (73) together, we obtain

$$\overline{R}_2(\omega) - R_2(\omega) \leq \max\left(2\Delta(\rho), \log\left(\frac{\pi e}{6}\right)+1\right).$$

By minimizing $\Delta(\rho)$ over $\rho > 0$, we obtain

$$\overline{R}_i(\omega) - R_i(\omega) \leq 12.772, \quad i = 1, 2.$$

Next we deal with the top maximum-sum-rate point. There are three cases:

Case 1: The top maximum-sum-rate point is below $L_0$. In this case, $\mathbb{E}\log(1+S_2) \leq \mathbb{E}\log\big(1+S_0/(1+S_1)\big)$, *i.e.*, the outer bound becomes a rectangle. Let both users use i.i.d. Gaussian signaling. Then the two users can achieve rates $\mathbb{E}\log(1+S_1)$ and $\mathbb{E}\log(1+S_2)$, respectively. Indeed, the condition $\mathbb{E}\log(1+S_2) \leq \mathbb{E}\log(1+S_0/(1+S_1))$ guarantees that receiver 1 can remove the signal of user 2 completely by treating its own signal as noise. Therefore, the gap between achievable scheme and outer bound is at most 1 bit for user 1 in this case. In fact, for this case, $0 \leq R_i \leq \mathbb{E}\log(1+S_i)$ $(i=1,2)$, is exactly the capacity region.

Case 2: The top maximum-sum-rate point is between $L_0$ and $L_b$. If the top maximum-sum-rate point is in set $\mathcal{M}$, then preceding analysis already covers this case. But it is possible that the top maximum-sum-rate point is between $L_0$ and $L_b$ and it is not in set $\mathcal{M}$. In other words, the part of curve $\mathcal{L}$ can be a line segment with slope steeper than $-1$ and the top maximum-sum-rate point is right on this segment. This situation can happen, for example, when $S_0$, $S_1$ and $S_2$ are all discrete random variables.

Denote the two ends of the line segment by $L_u$ and $L_d$, where $L_d$ is below the top maximum-sum-rate point. Suppose slope of the line segment is $-1/\omega_e$, then $L_u$'s coordinate is $\big(R_1(\omega_e), R_2(\omega_e)\big)$ and the coordinate of $L_d$ is

$$\left(1+\mathbb{E}\log(1+S_1) - \int_{\mathcal{B}^\circ(\omega_e)} \alpha(\gamma)\mathsf{d}\gamma \quad , \quad \mathbb{E}\log\left(1+\frac{S_0}{1+S_1}\right) + \int_{\mathcal{B}^\circ(\omega_e)} \beta(\gamma)\mathsf{d}\gamma\right) \tag{74}$$

where $\mathcal{B}^\circ(\omega) = \{\gamma \in [0,\infty) | \omega\beta(\gamma) > \alpha(\gamma)\}$. For notational convenience, we define $\mathcal{B}'(\omega) = \mathcal{B}(\omega)\backslash\mathcal{B}^\circ(\omega)$. Since both points are extreme points, they belong to $\mathcal{M}$. By preceding analysis, point $L_d$ can be achieved up to a gap of $12.772$ for each user. Therefore, it is sufficient to show the achievability result for the top maximum-sum-rate point because points between $L_d$ and the top maximum-sum-rate point can be shown by time-sharing argument. The point on the segment can be parametrized as

$$\begin{aligned}&\big(\widehat{R}_1(\delta), \widehat{R}_2(\delta)\big)\\ &= \left(1 + \mathbb{E}\log\left(1+S_1\right) - \int_{\mathcal{A}(\delta)} \alpha(\gamma)\mathsf{d}\gamma\ , \quad \mathbb{E}\log\left(1+\frac{S_0}{1+S_1}\right) + \int_{\mathcal{A}(\delta)} \beta(\gamma)\mathsf{d}\gamma\right)\end{aligned}$$

where

$$\mathcal{A}(\delta) = \mathcal{B}^o(\omega_e)\bigcup\Big(\mathcal{B}'(\omega_e)\cap[0,\delta)\Big).$$

It is easy to see that when $\delta$ varies from $0$ to $\infty$, $\big(\widehat{R}_1(\delta), \widehat{R}_2(\delta)\big)$ moves continuously from $L_d$ to $L_u$. Since the maximum-sum-rate point is between $L_u$ and $L_d$, it must be $\big(\widehat{R}(\delta^*), \widehat{R}_2(\delta^*)\big)$ for some $\delta^*$. Now, let $\mathcal{N}^* = \{n \geq 1 | \rho_n \in \mathcal{A}(\delta^*)\}$, where $\rho_n = \rho 2^{2(n-1)}$. Let user 1 generate its codebook according to i.i.d. unit CSCG distribution. Let user 2 generate the codebook for private message using the distribution of $X_{\mathcal{N}^*}$ and generate the codebook for common message using the distribution of $X_{\mathcal{N}^{*c}}$. Following the exactly same analysis as we did for the points in $\mathcal{M}$, we see that the top maximum-sum-rate point can be achieved up to a gap of $12.772$ for each user.

Case 3: The top maximum-sum-rate point is above $L_b$. In this case, the achievability result for $L_b$ holds by preceding analysis. That is following rate pair is achievable and it is at most $12.772$ away from the critical point for each user:

$$\begin{aligned}R_1(1) &= \mathcal{I}\left(\sqrt{S_1}W + \sqrt{S_0}X + U; W \middle| X_{\mathcal{N}^c(1)}, S_1, S_0\right)\\ R_2(1) &= \mathcal{I}\left(\sqrt{S_1}W + \sqrt{S_0}X + U; X_{\mathcal{N}^c(1)} \middle| S_1, S_0\right) + \mathcal{I}\left(\sqrt{S_2}X + V; X_{\mathcal{N}(1)} \middle| S_2\right).\end{aligned}$$

Since the outer bound between top maximum-sum-rate point and $L_b$ is a segment with slope -1, it suffices to investigate the achievability of points along the ray with slope -1 and starting from $\big(R_1(1), R_2(1)\big)$ till the intersection with the other constraint $R_2 = \mathbb{E}\log\left(1+S_2\right)$.

As in the study of layered erasure model, we split user 1 into two virtual users [27]. Let $W_1$ and $W_2$ be two independent CSCG random variables with $W_1 \sim \mathcal{CN}(0,\delta)$ and $W_2 \sim \mathcal{CN}(0, 1-\delta)$, where $\delta \in [0,1]$. Let user 1 generate the codebooks for its two virtual users using the distribution of $W_1$ and $W_2$, respectively. Receiver 1 first decode $U$, then decode $X_{\mathcal{N}^c}$ and finally decode $V$ in order. Therefore, following three rates are achievable at receiver 1

$$\begin{aligned} R_{1,1}(\delta) &= \mathcal{I}\left(\sqrt{S_1}\,(W_1+W_2)+\sqrt{S_0}\,X+U;U\Big|S_0,S_1\right) \\ R_{2c}(\delta) &= \mathcal{I}\left(\sqrt{S_1}\,(W_1+W_2)+\sqrt{S_0}\,X+U;X_{\mathcal{N}^c(1)}\Big|W_1,S_0,S_1\right) \\ R_{1,2}(\delta) &= \mathcal{I}\left(\sqrt{S_1}\,(W_1+W_2)+\sqrt{S_0}\,X+U;V\Big|X_{\mathcal{N}^c(1)},W_1,S_0,S_1\right)\ . \end{aligned}$$

Receiver 2 can achieve the private rate

$$R_{2p}(\delta) = \mathcal{I}\left(\sqrt{S_2}\,X+V;X_{\mathcal{N}(1)}\Big|S_2\right)\ .$$

By the chain rule, we have $R_{1,1}(\delta)+R_{1,2}(\delta)+R_{2c}(\delta)+R_{2p}(\delta) = R_1(1)+R_2(1)$. Therefore, we show that $\big(R_{1,1}(\delta)+R_{1,2}(\delta), R_{2c}(\delta)+R_{2p}(\delta)\big)$ is on the ray with slope -1 and starting from $\big(R_1(1),R_2(1)\big)$. That is the gap between the segment $\big(R_{1,1}(\delta)+R_{1,2}(\delta), R_{2c}(\delta)+R_{2p}(\delta)\big)$, $\delta \in [0,1]$, and the boundary of $H(1)$ is at most $12.772$.

Furthermore, we need to show that $R_{2c}(\delta)+R_{2p}(\delta)$ can go sufficiently close to line $R_2 = \mathbb{E}\log(1+S_2)$. Note that $R_{2c}(\delta)+R_{2p}(\delta)$ increases as $\delta$ increases, since larger $\delta$ implies that larger part of user 1's signal is removed before decoding the common message at receiver 1. Letting $\delta=1$,

$$R_{2c}(1)+R_{2p}(1) = \mathcal{I}\left(\sqrt{S_0}\,X+V;X_{\mathcal{N}^c(1)}\Big|S_0\right)+\mathcal{I}\left(\sqrt{S_2}\,X+V;X_{\mathcal{N}(1)}\Big|S_2\right)\ .$$

By Lemma 4,

$$\begin{aligned} R_{2c}(1)+R_{2p}(1)+2\Delta(\rho) &\geq \int_{\mathcal{B}^c(1)} \mathsf{P}\left(S_0\geq\gamma\right)\frac{\log e}{1+\gamma}\mathsf{d}\gamma + \int_{\mathcal{B}(1)} \mathsf{P}\left(S_2\geq\gamma\right)\frac{\log e}{1+\gamma}\mathsf{d}\gamma \\ &\geq \int_0^\infty \mathsf{P}\left(S_2\geq\gamma\right)\frac{\log e}{1+\gamma}\mathsf{d}\gamma \\ &= \mathbb{E}\log\left(1+S_2\right) \end{aligned}$$

where the last inequality is due to the fact $\mathsf{P}\left(S_0\geq\gamma\right)\geq\mathsf{P}\left(S_2\geq\gamma\right)$ whenever $\gamma\in\mathcal{B}^c(1)$. This shows that we can choose $\delta$ such that the top maximum-sum-rate point can be achieved up to $12.772$ for each user if the top maximum-sum-rate point is above $L_b$. This completes the proof of Theorem 1.

### C. Capacity Results for Some Special Cases

In the remainder of this section, we establish the capacity region for a few special cases which is not implied by Theorem 1.

*Theorem 3:* In case of one-sided stochastically strong interference, *i.e.*, $\mathsf{P}(S_0 \geq \gamma) \geq \mathsf{P}(S_2 \geq \gamma)$ for all $\gamma \geq 0$, the capacity region is

$$\left\{ (R_1, R_2): \begin{array}{r} 0 \leq R_1 \leq \mathbb{E}\log(1+S_1) \\ 0 \leq R_2 \leq \mathbb{E}\log(1+S_2) \\ R_1 + R_2 \leq \mathbb{E}\log(1+S_1+S_2) \end{array} \right\} \tag{75}$$

Theorem 3 directly follows the capacity results for general strong interference channel established in [24].

*Theorem 4:* In case of one-sided stochastically weak interference, *i.e.*, $\mathsf{P}(S_0 \geq \gamma) \leq \mathsf{P}(S_2 \geq \gamma)$ for all $\gamma \geq 0$, the sum-capacity of channel (1) is given by

$$C_{sum} = \mathbb{E}\log\left(1 + \frac{S_1}{1+S_0}\right) + \mathbb{E}\log(1+S_2) \tag{76}$$

The proof of Theorem 4 is relegated to Appendix IV. By (76), to achieve the sum-capacity, we need to let user 2 transmit at its full rate. Intuitively, the rate gain of the user 2 is always larger than the rate loss of user 1. In the proof of converse, we show that it is optimal for both users to use Gaussian signaling. This does not lead to the extra 1 bit compensation on user 1's rate as we did in proof of Theorem 1. Therefore, we can establish a tighter result in this case. It remains to see whether the extra $1$ bit in the third constraint of (3) can be removed in the general cases.

## VII. Conclusion

This work derives the first constant-gap result for the capacity region of the ergodic one-sided fading Gaussian interference channel with channel state information at the receiver but not at the transmitters. To achieve this, the new outer bound is obtained via investigating the trade-off between rate gain and rate loss of the two users. The achievability strategy is constructed by artificially layering one transmit signal. Both of the outer and inner bounds are motivated by the simpler and exact results for the corresponding layered erasure model.

# APPENDIX I
## PROOF OF LEMMA 2

Because signals are aligned at their respective least significant bit in a sum, one can write

$$\begin{aligned}
&\mathcal{H}\left(\boldsymbol{X}_1^{N_0}\middle|\boldsymbol{X}_1^{N_0}\oplus\widetilde{\boldsymbol{W}}_1^{N_1},\boldsymbol{T},\boldsymbol{N}\right)\\
&=\sum_{n_1=0}^{q}\sum_{n_0=0}^{q}\mathsf{P}\left(N_1=n_1\right)\mathsf{P}\left(N_0=n_0\right)\mathcal{H}\left(\boldsymbol{X}_1^{n_0}\middle|\boldsymbol{X}_1^{(n_0-n_1)^+},\boldsymbol{T}\right)\\
&=\sum_{n_1=0}^{q}\sum_{n_0=0}^{q}\sum_{l=1}^{q}\mathsf{P}\left(N_1=n_1\right)\mathsf{P}\left(N_0=n_0\right)1_{((n_0-n_1)^+<l\le n_0)}\mathcal{H}\left(X_l|\boldsymbol{X}_1^{l-1},\boldsymbol{T}\right) \qquad (77)\\
&=\sum_{l=1}^{q}\mathsf{P}\left(N_0-N_1<l\le N_0\right)\mathcal{H}\left(X_l|\boldsymbol{X}_1^{l-1},\boldsymbol{T}\right)\\
&=\sum_{l=1}^{q}\alpha_l\mathcal{H}\left(X_l|\boldsymbol{X}_1^{l-1},\boldsymbol{T}\right)
\end{aligned}$$

where (77) is due to the chain rule. Hence the proof of (25). Similarly,

$$\begin{aligned}
&\mathcal{H}\left(\boldsymbol{X}_1^{N_2}\middle|\boldsymbol{X}_1^{N_0}\oplus\widetilde{\boldsymbol{W}}_1^{N_1},\boldsymbol{T},\boldsymbol{N}\right)\\
&=\sum_{n_1,n_0,n_2}\mathsf{P}\left(N_1=n_1\right)\mathsf{P}\left(N_0=n_0,N_2=n_2\right)\mathcal{H}\left(\boldsymbol{X}_{(n_0-n_1)^++1}^{n_2}\middle|\boldsymbol{X}_1^{(n_0-n_1)^+},\boldsymbol{T}\right)\\
&=\sum_{n_1,n_0,n_2}\mathsf{P}\left(N_1=n_1\right)\mathsf{P}\left(N_0=n_0,N_2=n_2\right)\sum_{l=1}^{q}1_{((n_0-n_1)^+<l\le n_2)}\mathcal{H}\left(X_l|\boldsymbol{X}_1^{l-1},\boldsymbol{T}\right)\\
&=\sum_{l=1}^{q}\mathsf{P}\left(N_0-N_1<l\le N_2\right)\mathcal{H}\left(X_l|\boldsymbol{X}_1^{l-1},\boldsymbol{T}\right)
\end{aligned}$$

Furthermore,

$$\begin{aligned}
&\mathsf{P}\left(N_0-N_1<l\le N_2\right)\\
&=\mathsf{P}\left(N_2\ge l\right)-\mathsf{P}\left(N_0-N_1\ge l,N_2\ge l\right)\\
&=\sum_{n_1=0}^{q}\mathsf{P}\left(N_1=n_1\right)\left(\mathsf{P}\left(N_2\ge l\right)-\mathsf{P}\left(N_0-n_1\ge l,N_2\ge l\right)\right)\\
&=\sum_{n_1=0}^{q}\mathsf{P}\left(N_1=n_1\right)\left(\mathsf{P}\left(N_2\ge l\right)-\min(\mathsf{P}\left(N_0\ge n_1+l\right),\mathsf{P}\left(N_2\ge l\right))\right) \qquad (78)\\
&=\mathbb{E}\left[\mathsf{P}\left(N_2\ge l\right)-\mathsf{P}\left(N_0-N_1\ge l\middle|N_1\right)\right]^+\\
&=\beta_l
\end{aligned}$$

where (78) is due to the alignment between $N_0$ and $N_2$. Hence the proof of (26).

## Appendix II
## Proof of Lemma 3

Since $X$—$\widetilde{Z}$—$\widetilde{Y}$ is Markovian, we have

$$\begin{aligned}
&\mathcal{I}\left(\widetilde{Z};X\middle|\widetilde{Y},\boldsymbol{S},\boldsymbol{T}\right)\\
&=\mathcal{I}\left(\widetilde{Z};X\middle|\boldsymbol{S},\boldsymbol{T}\right)-\mathcal{I}\left(\widetilde{Y};X\middle|\boldsymbol{S},\boldsymbol{T}\right)\\
&=\log e\,\mathbb{E}\int_0^{S_0}\mathsf{mmse}\left[X|\gamma,\boldsymbol{T}\right]\mathsf{d}\gamma-\log e\mathbb{E}\int_0^{S_0/(S_1+1)}\mathsf{mmse}\left[X|\gamma,\boldsymbol{T}\right]\mathsf{d}\gamma \qquad (79)\\
&=\log e\,\mathbb{E}\int_{S_0/(S_1+1)}^{S_0}\mathsf{mmse}\left[X|\gamma,\boldsymbol{T}\right]\mathsf{d}\gamma\\
&=\log e\int_0^{\infty}\mathbb{E}1_{(S_0/(S_1+1)<\gamma\le S_0)}\mathsf{mmse}\left[X|\gamma,\boldsymbol{T}\right]\mathsf{d}\gamma\\
&=\log e\int_0^{\infty}\mathsf{P}\left(\frac{S_0}{S_1+1}<\gamma\le S_0\right)\mathsf{mmse}\left[X|\gamma,\boldsymbol{T}\right]\mathsf{d}\gamma\\
&=\log e\int_0^{\infty}\alpha(\gamma)\mathsf{mmse}\left[X|\gamma,\boldsymbol{T}\right]\mathsf{d}\gamma
\end{aligned}$$

where (79) is obtained by using the integral representation of mutual information via MMSE [28]. Hence the proof of (56).

Note that for any realization of $\boldsymbol{S}$, either $X$—$\overline{Z}$—$\widetilde{Y}$ or $X$—$\widetilde{Y}$—$\overline{Z}$ is Markovian. Furthermore, when it is the latter one, the mutual information is zero. Therefore, for every realization $\boldsymbol{s}$ of the states,

$$\mathcal{I}\left(\overline{Z};X\middle|\widetilde{Y},\boldsymbol{S}=\boldsymbol{s},\boldsymbol{T}\right)=\left(\mathcal{I}\left(\overline{Z};X\middle|\boldsymbol{S}=\boldsymbol{s},\boldsymbol{T}\right)-\mathcal{I}\left(\widetilde{Y};X\middle|\boldsymbol{S}=\boldsymbol{s},\boldsymbol{T}\right)\right)^{+}.$$

Thus,

$$\begin{aligned}
&\mathcal{I}\left(\overline{Z};X\middle|\widetilde{Y},\boldsymbol{S},\boldsymbol{T}\right)\\
&=\int\mathsf{P}_{S_1}(\mathsf{d}s_1)\int\mathsf{P}_{\Lambda}(\mathsf{d}\lambda)\left(\mathcal{I}\left(\overline{Z};X\middle|\boldsymbol{S}=\boldsymbol{s},\boldsymbol{T}\right)-\mathcal{I}\left(\widetilde{Y},X\middle|\boldsymbol{S}=\boldsymbol{s},\boldsymbol{T}\right)\right)^{+}\\
&=\log e\,\mathbb{E}\left[\int_0^{S_2}\mathsf{mmse}\left[X|\gamma,\boldsymbol{T}\right]\mathsf{d}\gamma-\int_0^{S_0/(S_1+1)}\mathsf{mmse}\left[X|\gamma,\boldsymbol{T}\right]\mathsf{d}\gamma\right]^{+}\\
&=\log e\,\mathbb{E}\int_0^{\infty}1_{(S_0/(S_1+1)<\gamma\le S_2)}\mathsf{mmse}\left[X|\gamma,\boldsymbol{T}\right]\mathsf{d}\gamma\\
&=\log e\int_0^{\infty}\mathsf{P}\left(S_0/(S_1+1)<\gamma\le S_2\right)\mathsf{mmse}\left[X|\gamma,\boldsymbol{T}\right]\mathsf{d}\gamma\,.
\end{aligned}$$

Furthermore,

$$\begin{aligned}
&\mathsf{P}\left(S_0/(S_1+1) < \gamma \le S_2\right) \\
&= \mathsf{P}\left(S_2 \ge \gamma\right) - \mathsf{P}\left(S_0/(S_1+1) \ge \gamma, S_2 \ge \gamma\right) \\
&= \mathsf{P}\left(S_2 \ge \gamma\right) - \mathbb{E}\min\left(\mathsf{P}\left(S_0/(S_1+1) \ge \gamma | S_1\right), \mathsf{P}\left(S_2 \ge \gamma\right)\right) \qquad (80) \\
&= \mathbb{E}\left[\mathsf{P}\left(S_2 \ge \gamma\right) - \mathsf{P}\left(S_0/(S_1+1) \ge \gamma | S_1\right)\right]^+ . \\
&= \beta(\gamma)
\end{aligned}$$

where (80) is due to the alignment between $S_0$ and $S_2$. Hence the proof of (57).

## Appendix III
## Proof of Lemma 4

Instead of computing the mutual information directly, the authors of [17] show that the system with input $X_{\mathcal{N}}$ and output $Z$ has an achievable rate at least the amount of RHS of (69). Hence (69) holds by coding theorem of point-to-point system [21].

To prove (70), we apply (69) to the sets $\mathcal{B}^c$ and $\mathcal{N}^c$, where $\mathcal{N}^c = \{1, 2, \dots\} \backslash \mathcal{N}$,

$$\int_{\mathcal{B}^c} \mathsf{P}\left(\Gamma \ge \gamma\right) \frac{\log e}{1+\gamma} \mathsf{d}\gamma - \Delta(\rho) \le \mathcal{I}\left(Z; X_{\mathcal{N}^c} \middle| \Gamma, \Theta'\right) .$$

It can be rewritten as

$$\begin{aligned}
\mathbb{E}\log\left(1+\Gamma\right) &- \int_{\mathcal{B}} \mathsf{P}\left(\Gamma \ge \gamma\right) \frac{\log e}{1+\gamma} \mathsf{d}\gamma - \Delta(\rho) \\
&\le \mathcal{I}\left(Z; X \middle| \Gamma\right) - \mathcal{I}\left(Z; X_{\mathcal{N}} | X_{\mathcal{N}^c}, \Gamma\right) \\
&= \mathcal{I}\left(Z; X \middle| \Gamma\right) - \mathcal{I}\left(\sqrt{\Gamma}\, X_{\mathcal{N}} + U_1; X_{\mathcal{N}} | \Gamma\right) . \qquad (81)
\end{aligned}$$

Note that $X$ has a uniform distribution on unit square. By [26, Eq. (10)], we have

$$\mathbb{E}\log\left(1+\Gamma\right) - \mathcal{I}\left(Z; X \middle| \Gamma\right) \le \log\left(\frac{\pi e}{6}\right) + 1 . \qquad (82)$$

Comparing (81) and (82), we have established (70).

# Appendix IV
# Proof of Theorem 4

For the achievability, we let both users generate the codebook with unit CSCG distribution. Assign rate $\mathbb{E}\log(1+S_2)$ to user 2. Receiver 1 decodes its own message by treating the signal from user 2 as noise. Therefore, following rate pair is achievable

$$R_1 = \mathbb{E}\log\left(1+\frac{S_1}{1+S_0}\right)$$
$$R_2 = \mathbb{E}\log(1+S_2)$$

which, in turn, achieves the sum-capacity.

For the converse, without changing the capacity region, we assume without loss of generality that $S_0$ and $S_2$ are driven by the same uniform random variable $\Lambda$ and $\Theta_0 \equiv \Theta_2 = 0$. Note that with this modification, we have $S_0 \le S_2$. By Fano's inequality

$$\begin{aligned} nR_1 + nR_2 - n\delta_n &\le \mathcal{I}\left(Y^n; W^n \middle| \boldsymbol{S}^n, \boldsymbol{\Theta}^n\right) + \mathcal{I}\left(Z^n; X^n \middle| \boldsymbol{S}^n, \boldsymbol{\Theta}^n\right) \\ &\le \mathcal{I}\left(Y^n; W^n, X^n \middle| \boldsymbol{S}^n, \boldsymbol{\Theta}^n\right) - \mathcal{I}\left(\{Y^n; X^n \middle| W^n, \boldsymbol{S}^n, \boldsymbol{\Theta}^n\right) \\ &\qquad + \mathcal{I}\left(Z^n; X^n \middle| \boldsymbol{S}^n, \boldsymbol{\Theta}^n\right) \\ &\le n\mathbb{E}\log(1+S_0+S_1) - \mathcal{I}\left(\{\sqrt{S_0}X+U\}^n; X^n \middle| \boldsymbol{S}_1^n\right) \\ &\qquad + \mathcal{I}\left(\{\sqrt{S_2}X+U\}^n; X^n \middle| \boldsymbol{S}^n\right) . \end{aligned} \tag{83}$$

Using "Marton-like" expansion as we did for the outer bound of Gaussian model (see the development of (60)), rewrite the last two terms as

$$\begin{aligned} &\mathcal{I}\left(\{\sqrt{S_2}X+U\}^n; X^n \middle| \boldsymbol{S}^n\right) - \mathcal{I}\left(\{\sqrt{S_0}X+U\}^n; X^n \middle| \boldsymbol{S}^n\right) \\ &\quad = \sum_{i=1}^n \left\{\mathcal{I}\left(\sqrt{S_{2i}}X_i+U_i; X^n \middle| \boldsymbol{T}_i, S_{2i}\right) - \mathcal{I}\left(\sqrt{S_{0i}}X_i+U_i; X^n \middle| \boldsymbol{T}_i, S_{0i}\right)\right\} \end{aligned}$$

where $\boldsymbol{T}_i = \left(\{\sqrt{S_2}X+U\}^{i-1}, \{\sqrt{S_0}X+U\}_{i+1}^n, \boldsymbol{S}^{i-1}, \boldsymbol{S}_{i+1}^n\right)$.

Note that $\left(X^{i-1}, X_{i+1}^n\right)$—$X_i$—$\left(\sqrt{S_{2i}}X_i+U_i, \sqrt{S_{0i}}X_i+U_i\right)$ is Markovian. Therefore the difference between the two mutual informations can be further rewritten as

$$\begin{aligned} &\sum_{i=1}^n \left\{\mathcal{I}\left(\sqrt{S_{2i}}X_i+U_i; X_i \middle| \boldsymbol{T}_i, S_{2i}\right) - \mathcal{I}\left(\sqrt{S_{0i}}X_i+U_i; X_i \middle| \boldsymbol{T}_i, S_{0i}\right)\right\} \\ &\quad = \sum_{i=1}^n \mathbb{E}\int_{S_0}^{S_2} \mathsf{mmse}\left[X_i|\gamma, \boldsymbol{T}_i\right] \mathrm{d}\gamma \end{aligned}$$

where we use the integral representation of mutual information via MMSE [28]. Moreover,

$$\mathsf{mmse}\left[X_i|\gamma, \boldsymbol{T}_i\right] \leq \frac{1}{1+\gamma} \quad \forall \gamma \geq 0.$$

Hence, we can establish that

$$\begin{aligned}\mathcal{I}\left(\{\sqrt{S_2}\,X+U\}^n; X^n\middle|\boldsymbol{S}^n\right) - \mathcal{I}\left(\{\sqrt{S_0}\,X+U\}^n; X^n\middle|\boldsymbol{S}^n\right) &\\ \leq n\mathbb{E}\int_{S_0}^{S_2} \frac{1}{1+\gamma}\mathsf{d}\gamma &\\ = n\mathbb{E}\log\left(1+S_2\right) - n\mathbb{E}\log\left(1+S_0\right)\ . & \qquad (84)\end{aligned}$$

Comparing (83) and (84) yields

$$nR_1 + nR_2 - n\delta_n \leq n\mathbb{E}\log\left(1+S_0+S_1\right) + n\mathbb{E}\log\left(1+S_2\right) - n\mathbb{E}\log\left(1+S_0\right)\ .$$

Letting $n \to \infty$, we have

$$\begin{aligned}R_1 + R_2 &\leq \mathbb{E}\log\left(1+S_0+S_1\right) + \mathbb{E}\log\left(1+S_2\right) - \mathbb{E}\log\left(1+S_0\right)\\ &= \mathbb{E}\log\left(1+\frac{S_1}{1+S_0}\right) + \mathbb{E}\log\left(1+S_2\right)\end{aligned}$$

which completes the proof.